\documentclass[cernpreprint,texmf,txfonts,USenglish,texlive=2011,subfigure=true,biblatex=false]{atlasdoc}
\usepackage{atlaspackage} 
\usepackage{atlasphysics}
\usepackage{mathrsfs}
\usepackage{epsfig}
\usepackage{url}
\usepackage[numbers,sort&compress]{natbib}

\PreprintIdNumber{CERN-PH-EP-2015-108}

\hypersetup{pdftitle={Measurements of the top quark branching ratios into channels with leptons and quarks with the ATLAS detector},pdfauthor={The ATLAS Collaboration}}

\AtlasTitle{Measurements of the top quark branching ratios into channels with leptons and quarks with the ATLAS detector}

\author{The ATLAS Collaboration}

\AtlasRefCode{TOPQ-2014-01}

\AtlasJournal{PRD}

\AtlasAbstract{ 
Measurements of the branching ratios of 
top quark decays into leptons 
and jets using events with $t \overline t$ (top antitop) 
pairs are reported. Events were 
recorded with the ATLAS detector at the LHC in {\it pp} collisions at a center-of-mass energy of $7\tev$. 
The collected data sample corresponds to
an integrated luminosity of 4.6~${\rm fb}^{-1}$. 
The measured top quark branching ratios agree with the Standard Model predictions within the measurement 
uncertainties of a few percent.
}

\begin{document}

\maketitle

\tableofcontents
\clearpage

\section{Introduction}
\label{sec:Introduction}

In the Standard Model (SM), 100\%  of the top quark decays contain a $W$ boson and a down-type quark. Measurements of 
 the ratio of top branching fractions 
$B(t\rightarrow W+b$-quark$)/B(t\rightarrow W+$down-type quark)~\cite{ratCMS} and 
of single top production~\cite{stAtlas, stCMS, stTev} have shown that more than 95\% of the decays are to a $W$ boson and a $b$-quark. In the SM the branching ratio to the different leptons is the same since the decay proceeds via a $W$ 
boson , but in models of new physics, e.g. supersymmetry (SUSY), final states with $\tau$ leptons can be enhanced
or suppressed~\cite{susy}; thus measuring the inclusive cross section using final states with
$\tau$ leptons can be a good probe for new physics. The measured values  of the top quark branching 
ratios will deviate with respect to the SM predictions if the data sample selected to extract $t \overline{t}$ events contains   
final states without two $W$ bosons.
Examples of processes that would cause deviations include events with a top quark decaying to charged Higgs boson or with SUSY 
particles decaying to the supersymmetric partner of the $\tau$ lepton ($\tilde{\tau}$).
Limits on the top quark branching ratio to a charged Higgs boson and a $b$-quark have
been published by the CDF~\cite{CDFHcs}, D0 ~\cite{D0Hcs1,D0Hcs2}, ATLAS~\cite{AtlasHcs1,AtlasHcs2} and CMS~\cite{CMSHltau} collaborations. Another example of a final state that can change
the observed branching ratios is the pair production of supersymmetric 
partners of the top quark ($\tilde{t}$) decaying into $b\nu_{\tau}\tilde{\tau}$ followed by 
the $\tilde{\tau}$ decay into a $\tau$ lepton and the gravitino, predicted by gauge-mediated  
SUSY breaking models~\cite{gmsb}.

This article presents the first direct measurement of 
the top quark semileptonic
and all-hadronic branching ratios. 
The branching ratios can be more sensitive probes of deviations from SM expectations
than measuring cross sections in different channels, because of cancellation 
of systematic uncertainties. The  large number of $t \overline{t}$ pairs
produced at the LHC provides an opportunity to measure 
top quark branching ratios with high precision. 
These top quark branching ratios are expected to be determined by the $W$ boson 
branching ratios, which have been measured 
at LEP~\cite{Wbr} to be in good agreement with the SM expectations~\cite{Wbrt}.  
Observing any deviation  would be an indication of 
non-SM processes contributing 
to final states dominated by $t \overline{t}$ production. 
This article also presents a measurement of the inclusive $t \overline{t}$ 
cross section using events with an isolated charged lepton ($\mu$ or $e$)
and a $\tau$ lepton decaying hadronically ($\tau_\mathrm{had}$). Previous measurements of the cross section
at $\sqrt{s}=7\tev$ in this channel have been published by the ATLAS and CMS collaborations~\cite{PLB,CMSltau}.

The analysis uses the full data sample, 4.6 fb$^{-1}$, collected by the ATLAS experiment at the LHC from 
$pp$ collisions 
at $\sqrt{s}=7\tev$ between March and November 2011. Kinematic selection criteria are applied  that require one or both of the top and antitop quarks to decay into a 
final state with one isolated lepton  and a jet.  
At least one jet in the event must
be tagged as originating from a $b$-quark ($b$-tag). Seven mutually exclusive final states
are used in this analysis: $e$+jets, $\mu$+jets, $ee$+jets, $\mu\mu$+jets, $e\mu$+jets, $e\tau_\mathrm{had}$+jets and $\mu\tau_\mathrm{had}$+jets.
Branching ratios for semileptonic and purely hadronic 
top quark decays are obtained by combining these seven final states assuming 
that only SM processes contribute to the background and the top branching ratios to leptons and jets add up to one.

\section{Analysis Overview}
\label{sec:Overview}
Data samples enriched with $t \overline{t}$ events are selected by means of criteria
that are designed to accept two $W$ bosons and at least one $b$-quark. In every event, 
either an electron or a muon is required, with the aim to select $W\rightarrow \ell \nu$, where $\ell$ stands for either 
$e$ or $\mu$.
  The $\ell$ may 
be produced directly in $W\rightarrow\ell\nu$ boson decays or indirectly in $W\rightarrow\tau\nu$ decays.
Separate event channels
are classified depending on the decay of a second $W$ boson: $W\rightarrow$ jets for $\ell$+jets, $W\rightarrow\ell\nu$ for 
$\ell\ell'$+jets, or $W\rightarrow\tau_\mathrm{had}\nu$ for $\ell\tau_\mathrm{had}$+jets. 
Since the analysis does not distinguish  electrons or muons that originate
from a $\tau$ lepton decay from those that come from direct $W\rightarrow e \nu$ and $W\rightarrow \mu \nu$ decays, both are included in the $W\rightarrow \ell \nu$ decays.
The branching ratios are measured by taking ratios of the number of $t \overline{t}$
events extracted from the three channels; thus an important aspect of the event 
selection is to use similar criteria for the object selection in all final states, so as to allow the cancellation of systematic uncertainties in the ratios. Another important criterion is to ensure 
that no event contributes to
more than one channel. The channel with the largest background and smallest number of signal events is that containing $\ell\tau_\mathrm{had}$+jets; thus the event
selection and analysis were optimized to reduce the uncertainty in that channel 
(see Sec.~\ref{sec:common_objects}).

The number of $t \overline t$ events in a given channel is extracted by fitting
background and signal templates to data distributions. 
The template shapes are fixed while
their normalizations are allowed to vary. The signal templates 
are derived from $t \overline{t}$ Monte Carlo (MC) simulation, which assumes that the top quark decays to a $W$ boson and a $b$-quark with a 100\% branching ratio. This assumption affects the shape of 
the signal templates, and if it is not valid for the selected data, the measured 
branching ratios will deviate from the SM prediction. 
The amount of background varies significantly in each channel. It is almost 
negligible in the $e\mu$+jets channel
and larger than the signal in the $\ell\tau_\mathrm{had}$+jets channels. 
In the $\ell$+jets 
channels, three invariant masses from two- and three-jet systems and a transverse mass distribution are fitted, as described in detail in Sec.~\ref{sec:l+jets}, while in the 
$\ell\ell'$+jets channels the dilepton effective mass distributions from two different  
missing transverse momentum ($\met$) regions are used (see Sec.~\ref{sec:l+l}). 
 Because of the much larger background, 
which originates from jets misidentified as $\tau$ leptons, a very different approach is taken
in the $\ell\tau_\mathrm{had}$+jets channel. 
Instead of fitting a kinematic distribution,
the quantity fitted is a boosted decision tree (BDT) output~\cite{BDT}, 
a multivariate discriminant that separates jets from $\tau$ leptons decaying to hadrons (see Sec.~\ref{sec:l+tau}).

The details of how the inclusive production cross section and branching ratios 
are derived from 
the number of  $t \overline t$ events obtained from each channel are discussed in
Sec.~\ref{sec:BR}. The systematic uncertainties of the measurements are estimated
by varying each source of systematic uncertainty by $\pm 1\sigma$ in templates derived from MC simulation and fitting all the distributions with the new templates 
(see Sec.~\ref{sec:systematics}). The final results are given in Sec.~\ref{sec:results}.

\section{ATLAS Detector }
\label{s:detector}

The ATLAS detector~\cite{ATLAS} at the LHC covers nearly the entire solid angle around the collision 
point.\footnotemark[1] \footnotetext[1]{ATLAS uses a right-handed coordinate system with its origin
at the nominal interaction point in the center of the detector and the $z$-axis along the
beam pipe. The $x$-axis points to the center of the LHC ring, and the $y$-axis points upwards.
The azimuthal angle $\phi$ is measured around 
the beam axis and the polar
angle $\theta$ is the angle from the beam axis. The pseudorapidity is defined as
$\eta=-\ln[\tan(\theta/2)]$. The distance $\Delta R$ in $\eta$--$\phi$ space is defined as 
$\Delta R=\sqrt{(\Delta \phi)^2+(\Delta \eta)^2}$. The transverse momentum and energy are defined as $\pt=p\sin\theta$ and $E_T=E\sin\theta$, respectively}
It consists of an inner tracking detector surrounded by a thin superconducting solenoid, 
electromagnetic (EM) and hadronic calorimeters, and an external muon spectrometer incorporating three 
large superconducting toroid magnet assemblies. The inner tracking detector provides tracking
information in a pseudorapidity range $|\eta|<2.5$. 
The liquid-argon (LAr) EM sampling calorimeters cover a range of $|\eta|<3.2$
with fine granularity. An iron/scintillator tile calorimeter provides hadronic energy
measurements in the central rapidity range ($|\eta|<1.7$). The endcap and forward regions
are instrumented with LAr calorimeters for both the EM and hadronic energy measurements
covering $|\eta|<4.9$. The muon spectrometer
provides precise tracking information in a range of $|\eta|<2.7$.

In 2011, ATLAS used a three-level trigger system to select events. The level-1 trigger is implemented in hardware
using a subset of detector information to reduce the event rate to less than 75 kHz. This is followed by two software-based
 trigger levels, namely level-2 and the event filter, which together reduce the event rate to 
about 300~Hz recorded for analysis.

\section{Data and Monte Carlo Samples}

The present measurements use collision data with a center-of-mass energy of
$\sqrt{s}= 7\tev$ taken in 2011 and selected with a single-electron or a single-muon trigger.
 Taking into account selection criteria for good data quality, the
total integrated luminosity for the analyzed data sample is
4.6 fb$^{-1}$. 

The $t\bar{t}$ signal is modeled using the POWHEG~\cite{powheg1,powheg2}  
 event generator, interfaced to PYTHIA6 (v6.421)~\cite{pythia} with the Perugia 2011C tune~\cite{P2011C} for showering and hadronization, setting the top quark mass 
to $172.5\gev$ and using the next-to-leading-order (NLO) 
parton distribution function (PDF) set  CTEQ66~\cite{CTEQ}. 
The $t\overline{t}$ production cross section  used in the simulation
is normalized to 177~pb as 
obtained from next-to-next-to-leading-order (NNLO) plus next-to-next-to-leading-logarithm (NNLL) calculations~\cite{Czakon}.

The calculation of the backgrounds uses MC simulations of $W/Z$ production with multiple jets (matrix elements for the jets production include light quarks, 
 $c$, $\bar c$, $c \bar c$, $b \bar b$), 
single-top-quark, and diboson 
($WW$, $WZ$, $ZZ$) events. Single-top-quark events were generated using
MC@NLO (v4.01)~\cite{MCNLO3} interfaced with HERWIG (v6.520)~\cite{herwig} and JIMMY (v4.31)~\cite{Jimmy} to model parton showering, hadronization, and the underlying-event using PDF set CT10~\cite{CT10}.
$W$+jets events with up to five partons and $Z$+jets events with 
$m(\ell^{+}\ell^{-})>40\gev$ and up to five partons were generated by ALPGEN~\cite{Alpgen} (v2.13) interfaced to HERWIG plus JIMMY  and the  CTEQ6L1~\cite{CTEQ6L1} PDF set.
The MLM matching scheme~\cite{MLM} of the ALPGEN generator is used to remove overlaps
between matrix-element and parton-shower products.  Diboson events 
were generated using HERWIG plus JIMMY and  the MRSTMcal PDF set~\cite{MRSTM}. Scale factors are applied to each process
to match next-to-leading-order predictions.
The $\tau$ decays are handled by TAUOLA \cite{tauola}.

All samples of simulated events include the effect of multiple $pp$ interactions in the same and neighboring bunch crossings (pile-up). On average, nine minimum-bias events are overlaid on all simulated events
to match the pile-up conditions in data. The average number of $pp$ collisions in a bunch crossing ($<$$\mu$$>$)
depends on the instantaneous luminosity, which increased over time; $<$$\mu$$>$ varied from 5 at the beginning of the
run period to approximately 18 at the end.
The events are reweighted in order to make the distribution of the average 
number of interactions per bunch crossing match the one observed in data.
All MC events are simulated with a detailed GEANT4-based detector simulation ~\cite{GEANT,ATLASSim}
and are reconstructed with the same algorithms as used in data.

\section{Event Selection}
\label{sec:common_objects}
Events are selected using a single-muon trigger with a $\pt$
threshold of $18\gev$ or a single-electron trigger with a $E_{\rm T}$ threshold of $20\gev$, rising to $22\gev$ during 
periods of high instantaneous luminosity. The $\pt$ and $E_{\rm T}$ criteria used in the further analysis guarantee a high and constant trigger efficiency.
The same triggers and reconstructed object definitions are applied to all channels.
\newline
\newline
Muon candidates are selected using tracks from the inner 
detector matched with tracks in the muon spectrometer~\cite{muons}. They are required to have $\pt>20\gev$ and $|\eta|<2.5$ 
and to satisfy criteria designed to reduce the muon misidentification probability.
The muon must have a longitudinal impact parameter ($z_0$) with respect to the primary vertex of less than 2 mm.  
In addition, to suppress muons from heavy-quark decays, muons must pass the isolation cuts:
the calorimeter energy in a cone of size $\Delta R = 0.2$ around the muon track must be less than $4\gev$,
and the scalar sum of the $\pt$ of the tracks reconstructed in the inner tracker in a cone of $\Delta R = 0.3$ around the
muon track must be less than $2.5\gev$. 
If a muon overlaps within a 
cone of ${\Delta}R = 0.4$ with an electron candidate or  with a jet, as defined below, 
it is not considered to be isolated.
\newline
\newline
Electron candidates are required to satisfy cuts on calorimeter and tracking variables to separate isolated electrons from jets~\cite{electrons}. Electrons must fall
into the region $|\eta_{\text{cluster}}|<2.47$, where $|\eta_{\text{cluster}}|$ is the pseudorapidity 
of the calorimeter energy cluster associated with the electron, excluding the transition region 
between the barrel and endcap calorimeters at $1.37<|\eta_{\text{cluster}}|<1.52$, 
and have  $E_{\rm T}>25\gev$.
The electrons must also pass an $E_{\rm T}$ isolation cut within a cone of ${\Delta}R = 0.2$  derived for 90\% efficiency along with a $\pt$ isolation cut within a cone of ${\Delta}R = 0.3$ derived for 90\% efficiency for prompt electrons from $Z\rightarrow e^+e^-$ events. 
The electron must have $z_0$ with respect to the primary vertex of less than 2 mm. 
Finally, if the electron lies within a cone of ${\Delta}R = 0.4$  around the muon or between $0.2<{\Delta}R \le 0.4$ around a jet as defined below, the object is considered to be a muon or a jet, respectively.
\newline
\newline
Jets are reconstructed from clustered energy deposits in the calorimeters 
using the anti-\emph{$k_t$}~\cite{antikt} algorithm with a radius parameter $R = 0.4$. Jets are required to have a transverse momentum 
$\pt>25\gev$ and to be in the pseudorapidity range $|\eta|<2.5$. The summed scalar $\pt$ of tracks associated with the jet and associated with the primary vertex is required to be at least 75\% of the summed $\pt$ of all tracks associated with the jet~\cite{JES}.
Any jet close to a good electron, as defined above, is considered to be an electron if it lies within a  cone of ${\Delta}R = 0.2$ around the electron. 
\newline
\newline
Missing transverse momentum 
 ($\met$) is the magnitude of the vector sum of the $x$ and $y$ components of the cluster energy in the calorimeters. Each cluster is calibrated according to which type of high-$\pt$ object 
 it is matched to, either electrons, jets, muons or photons.
\newline
\newline
Jets containing $b$-hadrons ($b$-jets) are identified ($b$-tagged) with
a multivariate discriminant that exploits the long lifetimes,
high masses and high decay multiplicities of $b$-hadrons. It makes use of
track impact parameters and reconstructed secondary vertices. An operating point
corresponding to an average efficiency of 70\% 
and an average mistag rate for light-quark jets of 0.8\% is used~\cite{MV1}.
\newline
\newline
$\tau$ candidates are reconstructed using calorimeter jets as seeds. These seed jets are
calibrated with the local calibration (LC) scheme~\cite{LC1,LC2}.  
The $\tau$ candidate must have $E_{\rm T}^{\tau}>20\gev$, $|\eta_{\tau}|<2.3$, and only one
track with $\pt>4\gev$ associated with the $\tau$ candidate (77\% of hadronic $\tau$ decays
have only one track). The charge of the $\tau$ candidate is given by the charge of the associated track. 
Candidates with higher track multiplicity are not used as they do not 
improve the precision of the measurement because of much larger associated systematic 
uncertainties.
The analysis makes use of a BDT for $\tau$ identification, a cut-based multivariate algorithm that 
optimizes signal and background separation~\cite{BDT}.  

The $\tau$ candidates that overlap  within ${\Delta}R<0.4$ of a $b$-tagged jet, a loose muon,~\footnotemark[2] \footnotetext[2]{Loose muons are selected with all requirements described in Sec.~\ref{sec:common_objects}
for good muons, except $\pt^{\mu}>4\gev$ and no isolation requirements are applied.} or an
 electron,~\footnotemark[3] \footnotetext[3]{ These electrons are selected with all requirements described in Sec.~\ref{sec:common_objects} for good electrons, but electrons with $E_{\rm T}>20\gev$ are considered.} are rejected and kept as jets or electrons.
To remove the remaining electrons misidentified as $\tau$ candidates a medium BDT (BDT$_e$)
electron veto is applied. BDT$_e$ is a BDT trained to distinguish electrons and $\tau$ leptons
using a $Z\rightarrow\tau\tau$ MC sample as signal and a $Z\rightarrow\ell\ell$ MC sample as background. 
The BDT$_e$ uses four variables, the two most powerful being the ratio
 of high-threshold to low-threshold track hits in the transition
radiator and the ratio of energy deposited in the EM calorimeter to
the total energy deposited in the calorimeter.
The medium working point corresponds to 85\% efficiency
for $Z\rightarrow\tau\tau$, Ref.~\cite{taurec}.   
The additional rejection factor for electrons after removing isolated electrons that 
overlap with $\tau$ candidates is 60.
In addition, a muon veto 
that compares the track momentum in $\tau$ candidates with the energy
deposited in the electromagnetic calorimeter is required to further reduce 
the muon background. It is tuned to 96\% efficiency on signal (62\% on background 
after overlap removal). 
A BDT to reject hadronic jets faking $\tau$ leptons, BDT$_j$,
is trained with $\tau$ leptons from a $Z\rightarrow\tau\tau$ MC sample as signal and jets from data, 
selected from events with at least 
two jets, as background. The BDT$_j$ uses eight 
    variables, the most sensitive is the fraction of energy deposited in
    the region $\Delta R<0.1$ with respect to all energy deposited in
    the region $\Delta R<0.2$ around the $\tau$ candidate.
Details of the BDT$_e$ and BDT$_j$ input variables and performance are given in Ref.~\cite{taurec}.
\newline
~

The event selection requirements common to all channels are 
a primary vertex with at least five associated tracks
with $\pt>400\mev$, at least one isolated high-$\pt$ muon ($\pt>20\gev$) and/or isolated high-$\pt$ electron ($\pt>25\gev$),
at least two jets with $\pt>25\gev$, and at least one of them tagged as
a $b$-jet.
In addition, there are requirements specific to each channel.
For the $\ell$+jets channels  the isolated-muon $\pt$ threshold is raised from 
$20\gev$ to $25\gev$ to reduce the multijet background and exactly one isolated $\ell$ is required. The minimum number of jets with $\pt>$ 25 GeV is raised to four.
Events with $\tau$ candidates are removed.  Removing events with $\tau$ candidates 
from the $\ell$+jets channel results in an efficiency loss of 8.5\%.
For the $\ell\ell'$+jets channels, events are required to
have exactly two isolated $\ell$ with 
opposite-sign charges and  $\met>30\gev$. For the $\ell\tau_\mathrm{had}$+jets channels,
exactly one isolated $\ell$, $\met>30\gev$, and at least one $\tau$ candidate,  are required. In addition the $\ell$ and the $\tau$ candidate 
must have opposite charge. 
The $\tau$ candidates that do not satisfy these requirements are kept as jets. The  
thresholds for lepton $\pt$, jet $\pt$ and $\met$ were optimized for the $\ell\tau_\mathrm{had}$+jets channel for maximum signal significance by means of a search in parameter space.

\section{Single-lepton + jets channel}
\label{sec:l+jets}
Three different classes of events 
contribute as a background to the $t \overline{t}\rightarrow \ell$+jets channel: 
\begin{enumerate}
\item events with one isolated $\ell$ originating from processes with one true lepton ($W$ boson decay);
\item events with one jet misidentified as an isolated lepton and no other isolated lepton reconstructed;
\item events with one isolated lepton originating from processes with multiple true leptons but only one isolated lepton reconstructed.
\end{enumerate}

The number of $t \overline{t}\rightarrow \ell$+jets  events is extracted 
by fitting distributions of four invariant mass variables with templates for signal and 
backgrounds.
The following mass variables provide good discrimination between signal and background:

\begin{enumerate}
\item {\bf $m_{jj}$}: invariant mass of the two highest-$\pt$ jets
not designated as $b$-jets;
\item {\bf $m_{b1jj}$}: invariant mass of the leading $b$-jet
and the jets used to calculate $m_{jj}$;
\item {\bf $m_{b2jj}$}: invariant mass of the subleading $b$-jet
and the jets used to calculate $m_{jj}$;
\item {\bf $m_{\rm T}$}: transverse mass of $\ell$ and the $\met$, $m_{\rm T}(\ell,\met)=\sqrt{(E_{\rm T}^{\ell}+\met)^2-(p_x^{\ell}+E^{\rm miss}_x)^2-(p_y^{\ell}+E^{\rm miss}_y)^2}$.
\end{enumerate}

If an event has only one jet tagged as a $b$-jet, the highest-$\pt$ jet that is not tagged is assumed to 
be a second $b$-jet.
A few observations motivate the choice of mass distributions for the fit. 
The presence of a $W$ boson decaying to a pair of quarks leads to a $m_{jj}$ distribution that peaks at the $W$ boson mass.
The presence of a top quark decaying to $W(\rightarrow qq)+b$ will produce  $m_{b1jj}$ and $m_{b2jj}$ distributions that 
peak at the top quark mass.  
The presence of  a $W$ boson decaying to $\ell+\nu$
manifests itself as a Jacobian peak in the $m_{\rm T}$ distribution 
when there are no additional high-$\pt$ neutrinos in the event.

\subsection{Background templates}
\label{sec:bckljets}
The main backgrounds in the $\ell$+jets channel are from $W(\rightarrow\ell\nu)$+jets and other
$t \overline{t}$ final states. There are also smaller contributions from single top,
$Z(\rightarrow\ell\ell)$+jets (with one lepton not identified) 
and multijet processes with one jet misidentified
as a lepton.  Background templates are derived from the MC simulations in all cases except 
multijet processes. The multijet background is very difficult to simulate due 
to the need for a very large sample  
and the fact that MC models do not reproduce that  background well.
Instead it is derived from a control data sample with nonisolated electrons and muons,  
keeping all other selection criteria the same.
The distributions of a small expected contribution from $t \overline{t}$ is subtracted from the multijet control sample.

\begin{figure}[!hbt]
\begin{tabular}{cc}
\epsfig{figure=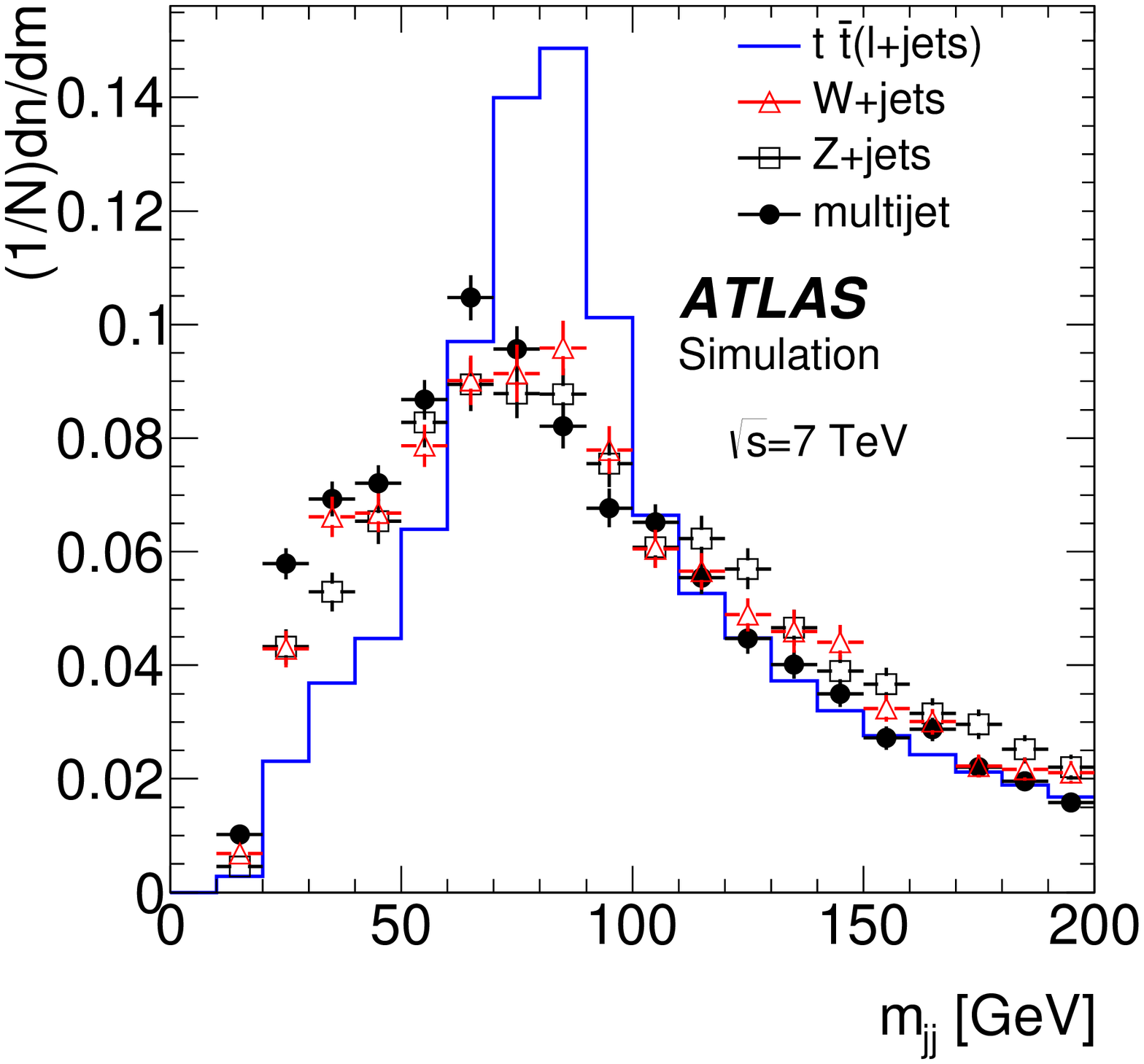,width=7.5cm} & 
\epsfig{figure=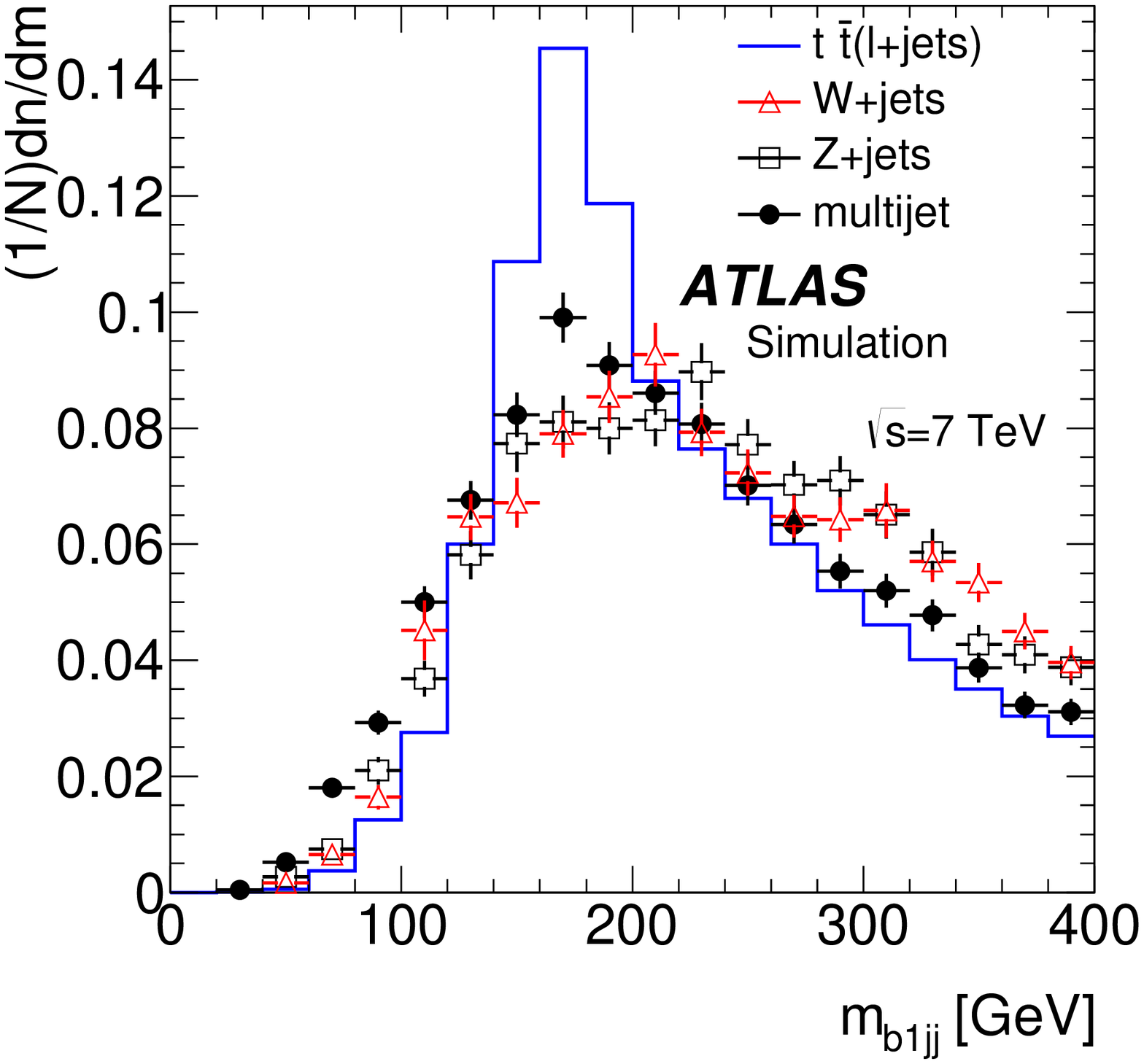,width=7.5cm} \\
(a) & (b) \\
\epsfig{figure=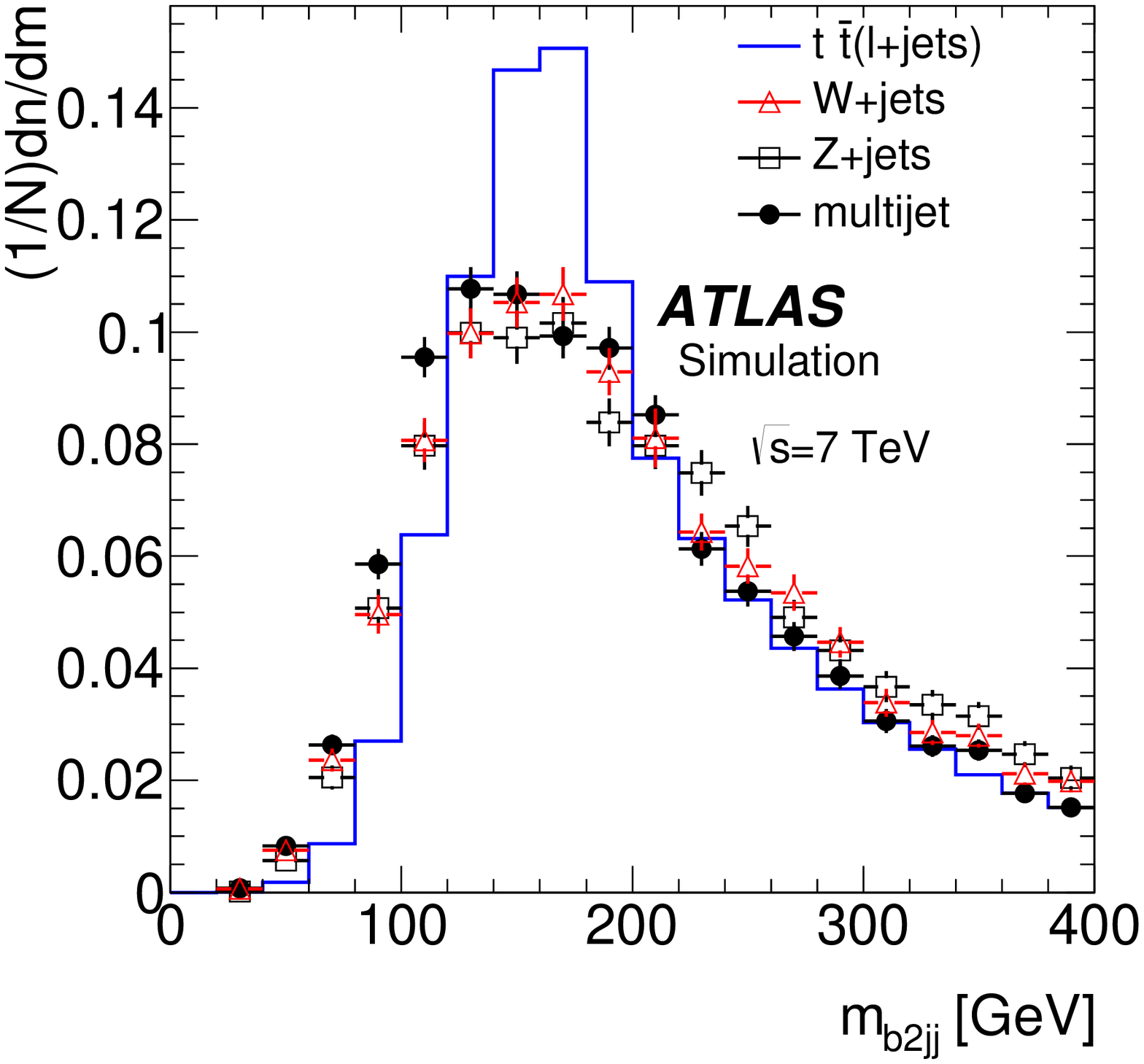,width=7.5cm} & 
\epsfig{figure=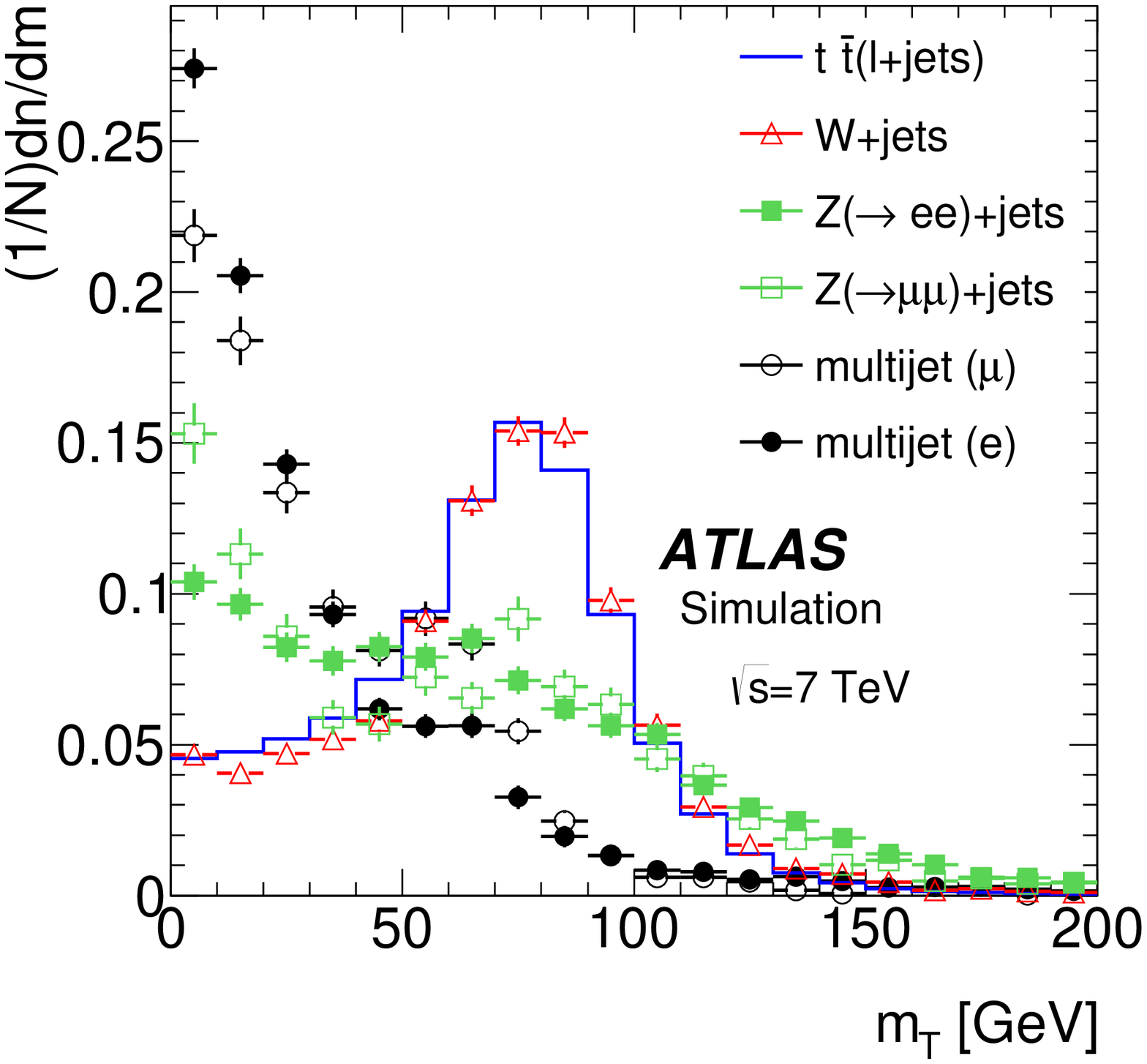,width=7.5cm} \\
(c) & (d) \\
\end{tabular}
\caption{\label{fig:m_qcd_z_w1} (a)  Invariant mass of two highest-$\pt$ jets not designated as $b$-jets ($m_{jj}$),   (b) and (c) invariant masses of jets designated as $b$-jets and the jets used for $m_{jj}$, ($m_{b1jj}$) and 
($m_{b2jj}$), where $b1$ stands for the leading $b$-jet and $b2$ for the subleading $b$-jet, and (d)  transverse mass of lepton and $\met$ ($m_{\rm T}$). The distributions have been  normalized and show distributions  for $t \overline{t}\rightarrow \ell$+jets,
 $Z(\rightarrow \ell \ell)$+jets,
 $W(\rightarrow\ell\nu)$+jets MC events and multijet events populating the $\ell$+jets channels. 
 The $e$ and $\mu$ channels have been merged together in the $m_{jj}$, $m_{b1jj}$ and $m_{b2jj}$ 
distributions. They are kept separate in the $m_{\rm T}$ distributions except for $t \overline{t}$
 and $W$+jets.
 Events are required to have exactly one 
 isolated $e$ or $\mu$, $\met>30\gev$, at least four jets, and at least one $b$-tagged jet.}
\end{figure}

Figure~\ref{fig:m_qcd_z_w1}
 shows the $m_{jj}$, $m_{b1jj}$, $m_{b2jj}$ and $m_{\rm T}$ distributions  predicted by MC simulation and normalized to unity for $W$+jets, $Z$+jets, and  $t \overline{t}\rightarrow\ell$+jets events. It also shows these distributions for multijet events derived from the control data
sample. The distributions from other $t \overline{t}$ channels are not shown 
as that background is normalized following the MC prediction of the ratio to the number of  $t \overline{t}\rightarrow\ell$+jets events.  
 The figure demonstrates that the shape of all the invariant 
mass distributions from jets  
are quite distinct for  $t \overline{t}\rightarrow \ell$+jets while there
is very little difference between the various backgrounds.
The distributions for $t \overline{t}\rightarrow\ell$+jets events
show that they include top quarks decaying to $b$+$W$ with the $W$ boson decaying to jets.
On the other hand, the $m_{\rm T}$ distributions show that they include a $W$ boson
decaying leptonically in both the $t \overline{t}\rightarrow\ell$+jets and $W$+jets channels
but cannot discriminate between them. They do show a clear separation between final states with one $W$ boson decaying leptonically 
and those with little intrinsic $\met$ ($Z$+jets and multijets).

\begin{figure}[!hbt]
\begin{tabular}{ccc}
\epsfig{figure=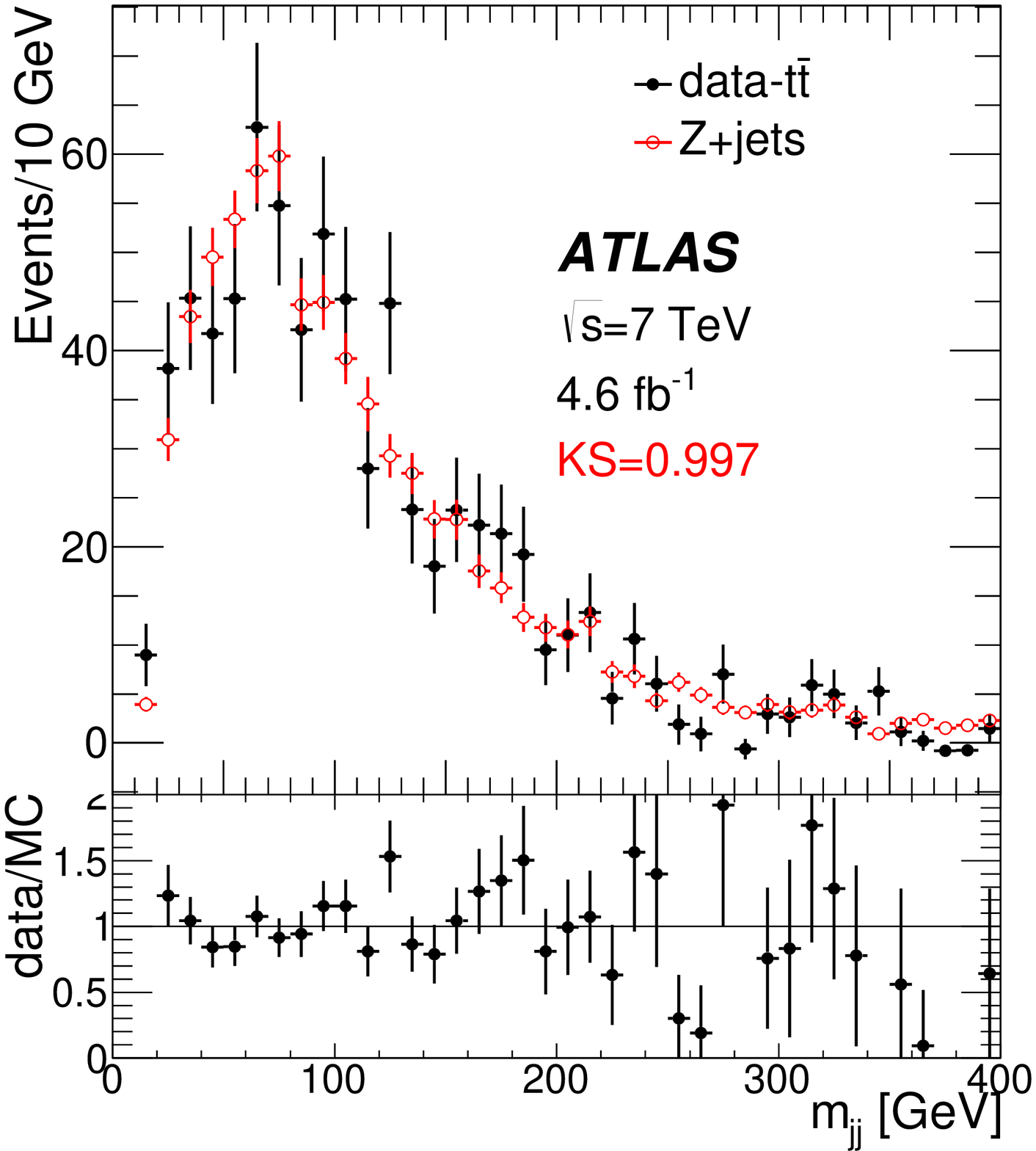,width=5.0cm, height=5.0cm} & 
\epsfig{figure=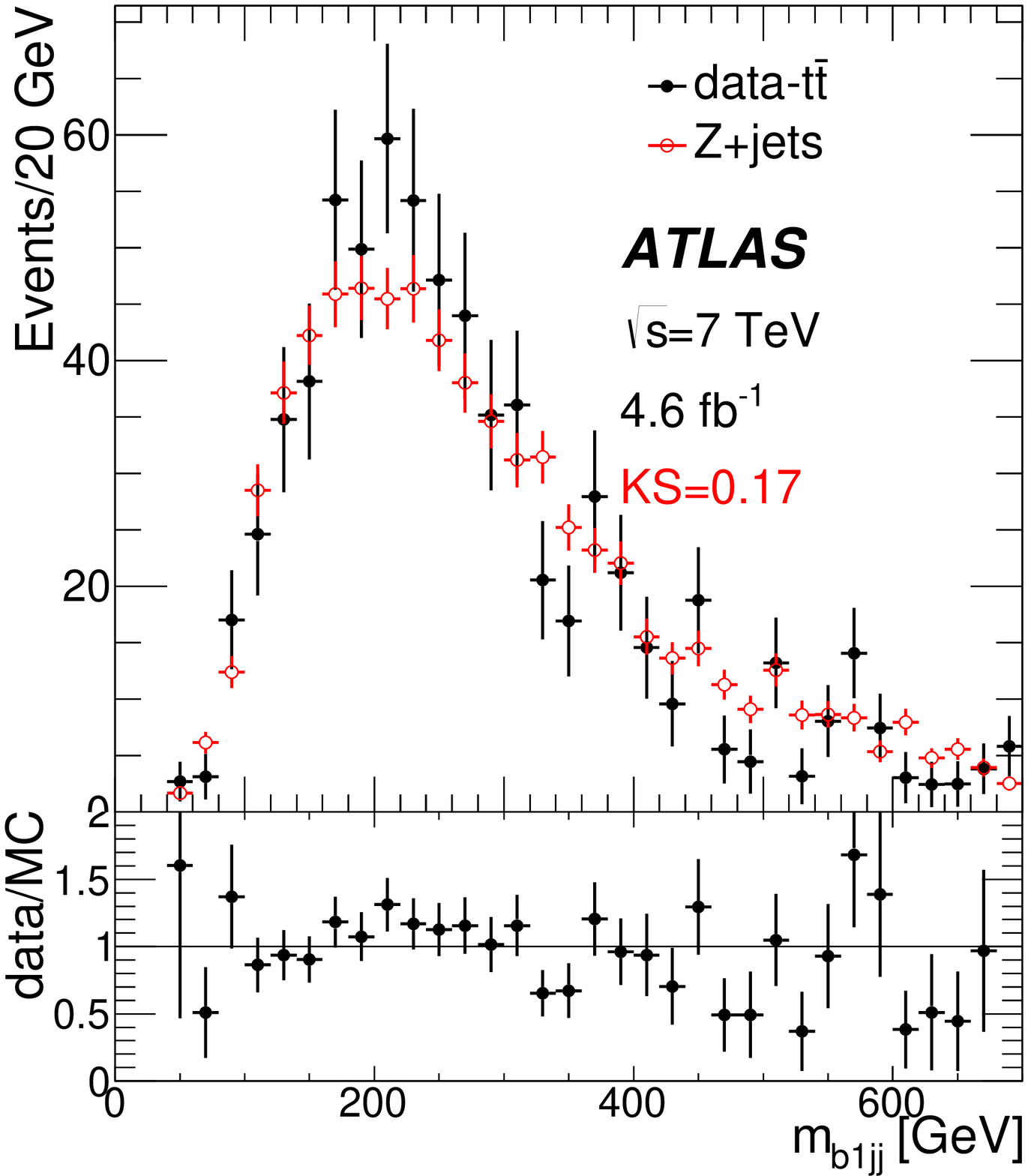,width=5.0cm, height=5.0cm} & 
\epsfig{figure=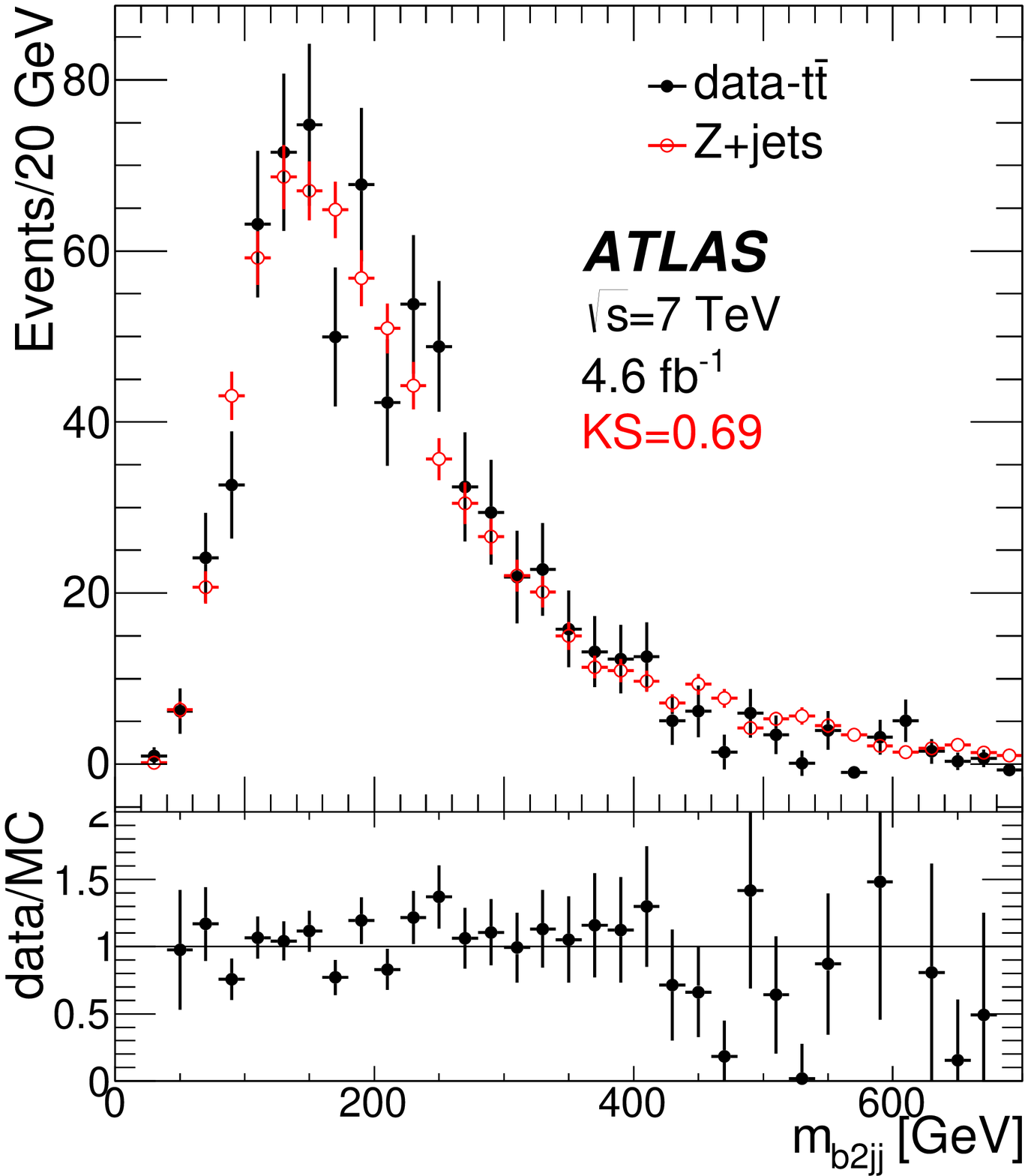,width=5.0cm, height=5.0cm} \\
(a) & (b) & (c) \\ 
\end{tabular}
\caption{\label{fig:Z} (a)  Invariant mass of two highest-$\pt$ jets not designated as $b$-jets ($m_{jj}$),   (b) and (c) invariant masses of jets designated as $b$-jets and the jets used for $m_{jj}$, ($m_{b1jj}$) and 
($m_{b2jj}$), where $b1$ stands for the leading $b$-jet and $b2$ for the subleading $b$-jet, and (d)  transverse mass of lepton and $\met$ ($m_{\rm T}$).  The distributions show ALPGEN MC for a control sample of 
$Z(\rightarrow\ell\ell)$+jets events selected by requiring $70<m_{\ell\ell}<110\gev$ ($m_{\ell\ell}$ is the invariant mass of the two leptons), $\met>30\gev$, at least four jets, and at least one of them $b$-tagged, compared to the data after subtracting the expected $t \bar t$ contribution. KS is the value of the Kolmogorov-Smirnov goodness-of-fit test.}
\end{figure}

The background templates for $Z$+jets events from MC simulation are checked with $Z$+jets events from data
by selecting events with two identified leptons and requiring the 
dilepton mass to be near the $Z$ mass. Events are required to have 
two oppositely charged leptons 
($\pt^e>25$ GeV and $\pt^{\mu}>20\gev$), $70\gev<m_{\ell\ell}<110\gev$, $\met>30\gev$, and the same jet selections as for the $\ell$+jets signal. 
The only significant background
 in the control data sample is from the $t \bar t\rightarrow \ell\ell'$+jets channel.
Figure~\ref{fig:Z} shows the $m_{jj}$, $m_{b1jj}$ and $m_{b2jj}$ distributions 
after merging $ee$ and $\mu\mu$ events for ALPGEN $Z$+jets MC simulation and the data after applying scale factors (SF) based on comparing data and simulation as a function of the $Z$ boson $\pt$ and the jet multiplicity. The 
small expected $t \bar t$ 
contribution is subtracted from the data distributions. 
The Kolmogorov-Smirnov goodness-of-fit test (KS) value in each plot indicates how
well the shape of the data distribution is described by the ALPGEN MC simulation.
\footnotemark[4] \footnotetext[4]{KS is calculated with the function supplied by ROOT for comparing
the compatibility of two histograms~\cite{ROOT}.}
Since there is no noticeable difference between the shapes
of the $W$+jets and $Z$+jets templates, as shown in Fig.~\ref{fig:m_qcd_z_w1}, one can conclude 
that both MC templates
can reproduce reasonably well the distributions expected in the data. The number 
of selected $Z$+jets events is also predicted well by the simulation.

\subsection{Fits to mass distributions}
\label{sec:ljets_fits}
As shown in Sec.~\ref{sec:bckljets} the three invariant masses constructed from jets do not discriminate between the various backgrounds, while the signal from $t \overline{t}$
is quite distinct. The only distribution that is different for each background is the
transverse mass. In particular, the transverse mass clearly distinguishes final states with intrinsic $\met$,
 i.e. those with a $W$ boson decaying to a lepton and neutrino, from those where $\met$ is due to mismeasurements.
The dominant processes without sizable intrinsic $\met$ are multijet and $Z$+jets. 
The transverse
mass distributions for those two processes are different. However, they contribute 
little in
$m_{\rm T}>40\gev$ so most of the separation comes from the region below 40$\gev$. 
As shown in Fig.~\ref{fig:Z}, the ALPGEN $Z$+jets simulation predicts the shape and the number of $Z$+jets events well,
so the choice is made to normalize the number of $Z$+jets events
to that predicted by the simulation. 
The number of single top events is similarly normalized from MC 
simulation. The amount of multijet background is obtained from the fit to the data using the templates derived from
nonisolated lepton samples. 
The other free parameters are the total number of $W$+jets events and the total number of $t \overline{t}$ events. The 
fractional contributions for the various $t \overline{t}$ channels are obtained using MC events.  
To ensure that events are not used more than once, two sets of data are fitted: 
$\met<30\gev$ (set 1) and $\met>30\gev$ (set 2).
Set 1 is used to fit the $m_{\rm T}$ distributions and helps determine the multijet background. 
Set 2 is used to fit the three jet mass distributions. 
Both sets are fit simultaneously with three parameters:
the total number of multijet events, the total number of $W$+jets events and the total number of $t \overline{t}$ events. 
 
The variables $m_{b1jj}$ and $m_{b2jj}$ are strongly correlated with $m_{jj}$. To exploit the fact that
the correlations are very different in $t \overline{t}$ and the background, the fits are done simultaneously in $6\times 6 \times 6$ bins of $m_{jj}$ ,$m_{b1jj}$ and $m_{b2jj}$ for a total of 216 bins.
Of those, 30 bins have zero events since they are kinematically not possible.  
The ranges and bin sizes are chosen so that all bins used for fitting are populated by more than 
10 events. That limits the range of $m_{\rm T}$ to $m_{\rm T}<120\gev$, $m_{jj}$
to $m_{jj}<250\gev$, $m_{b1jj}$ to $m_{b1jj}<450\gev$, and $m_{b2jj}$ to $m_{b2jj}<450\gev$.

\begin{figure}[!hbt]
\begin{tabular}{cc}
\epsfig{figure=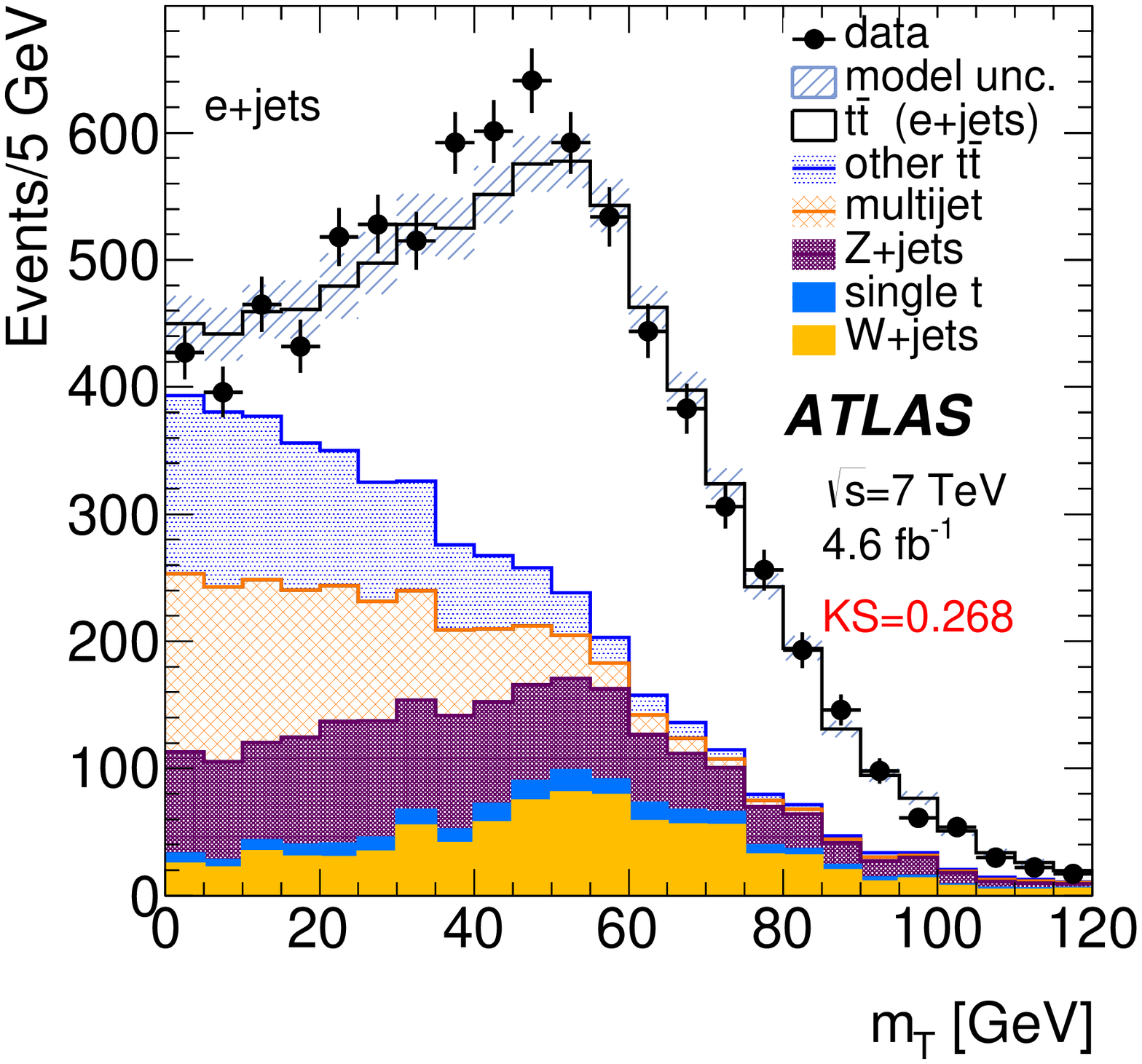,width=7.5cm} & 
\epsfig{figure=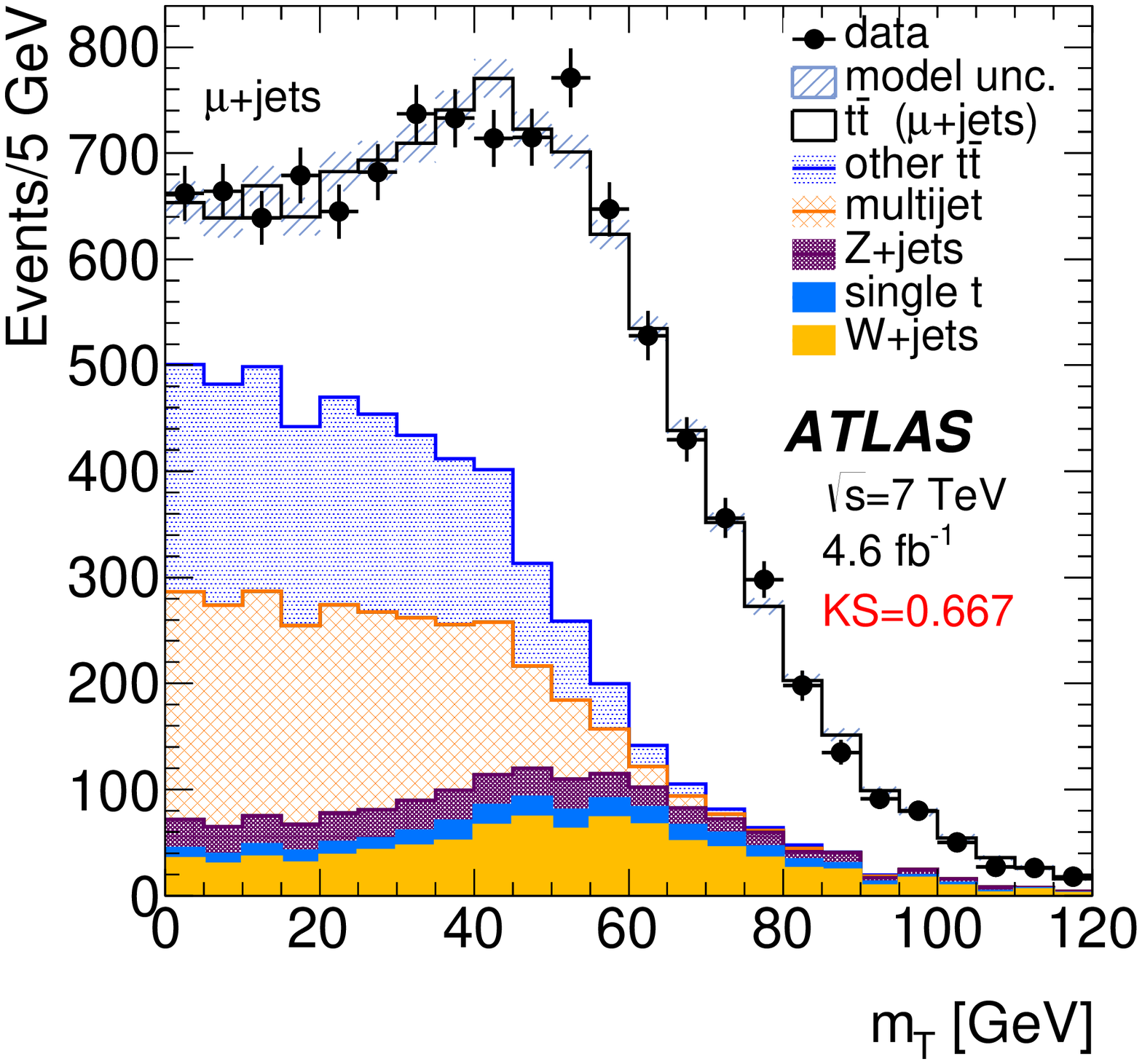,width=7.5cm} \\
\end{tabular}
\caption{\label{fig:ljets_Mt} Transverse mass of lepton and $\met$ ($m_{\rm T}$) distributions used in the fits.
Events are required to have exactly 
one isolated $e$ or $\mu$, $\met<30\gev$, at least four jets, and at least one $b$-tag.
The model uncertainty (model unc.) is the sum in quadrature of the statistical uncertainties of the templates used in the fits. KS is the value of the Kolmogorov-Smirnov goodness-of-fit test.}
\end{figure}

\begin{figure}[!hbt]
\begin{tabular}{cc}
\epsfig{figure=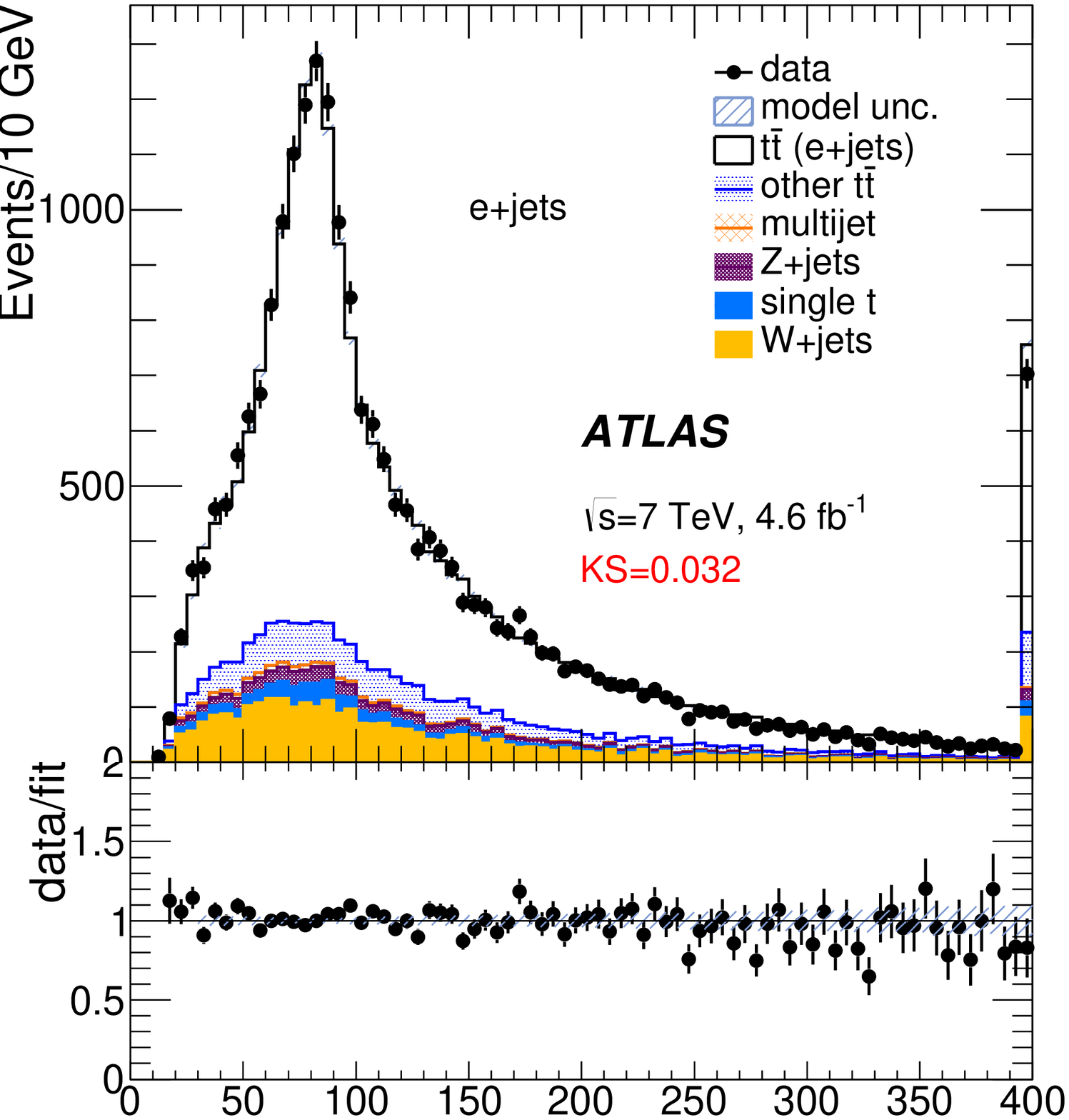,width=7.5cm, height=6.5cm} & 
\epsfig{figure=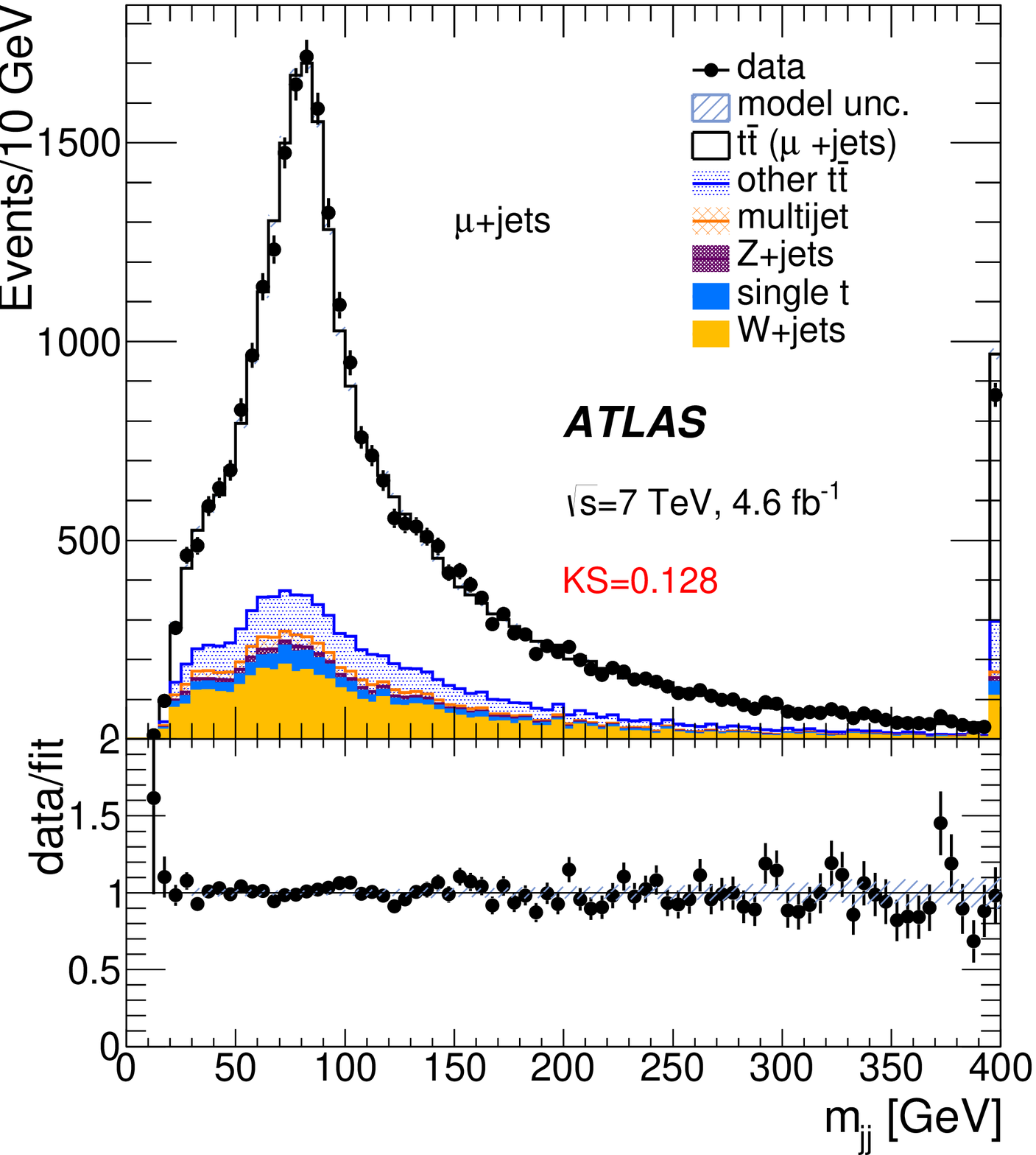,width=7.5cm, height=6.5cm} \\
(a) $e$+jets& (b) $\mu$+jets\\
\epsfig{figure=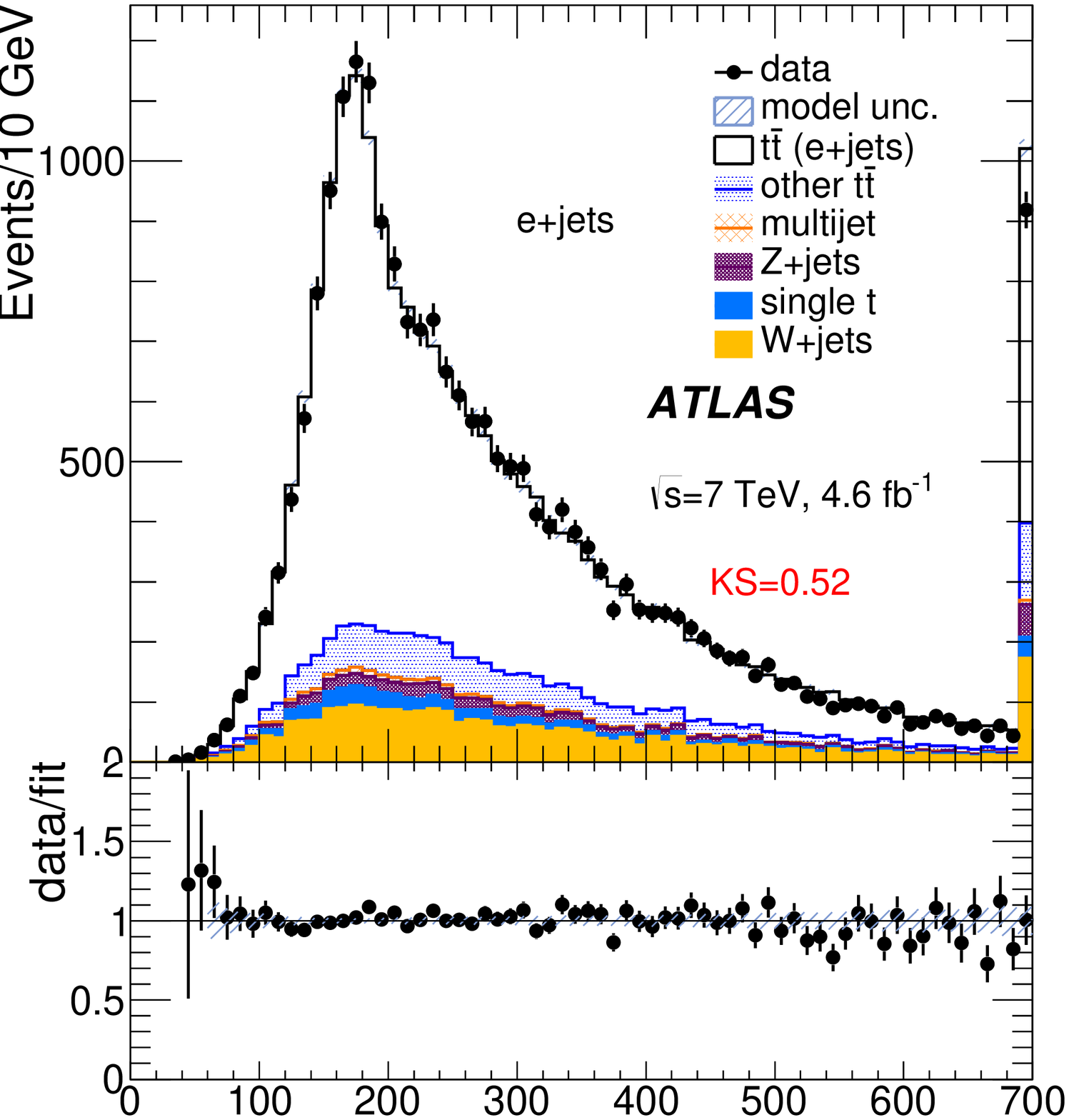,width=7.5cm, height=6.5cm} & 
\epsfig{figure=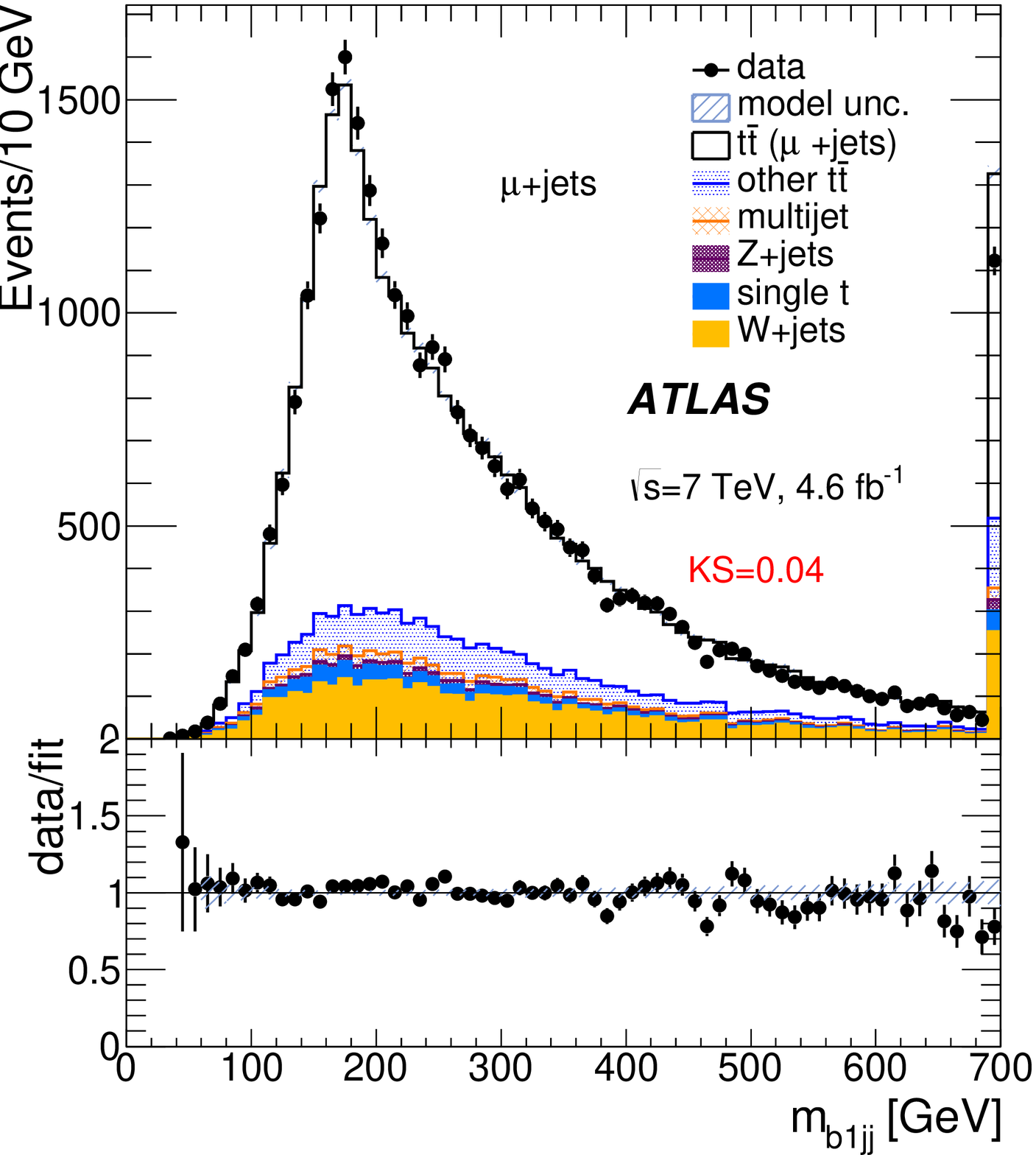,width=7.5cm, height=6.5cm} \\ 
(c) $e$+jets & (d) $\mu$+jets\\
\epsfig{figure=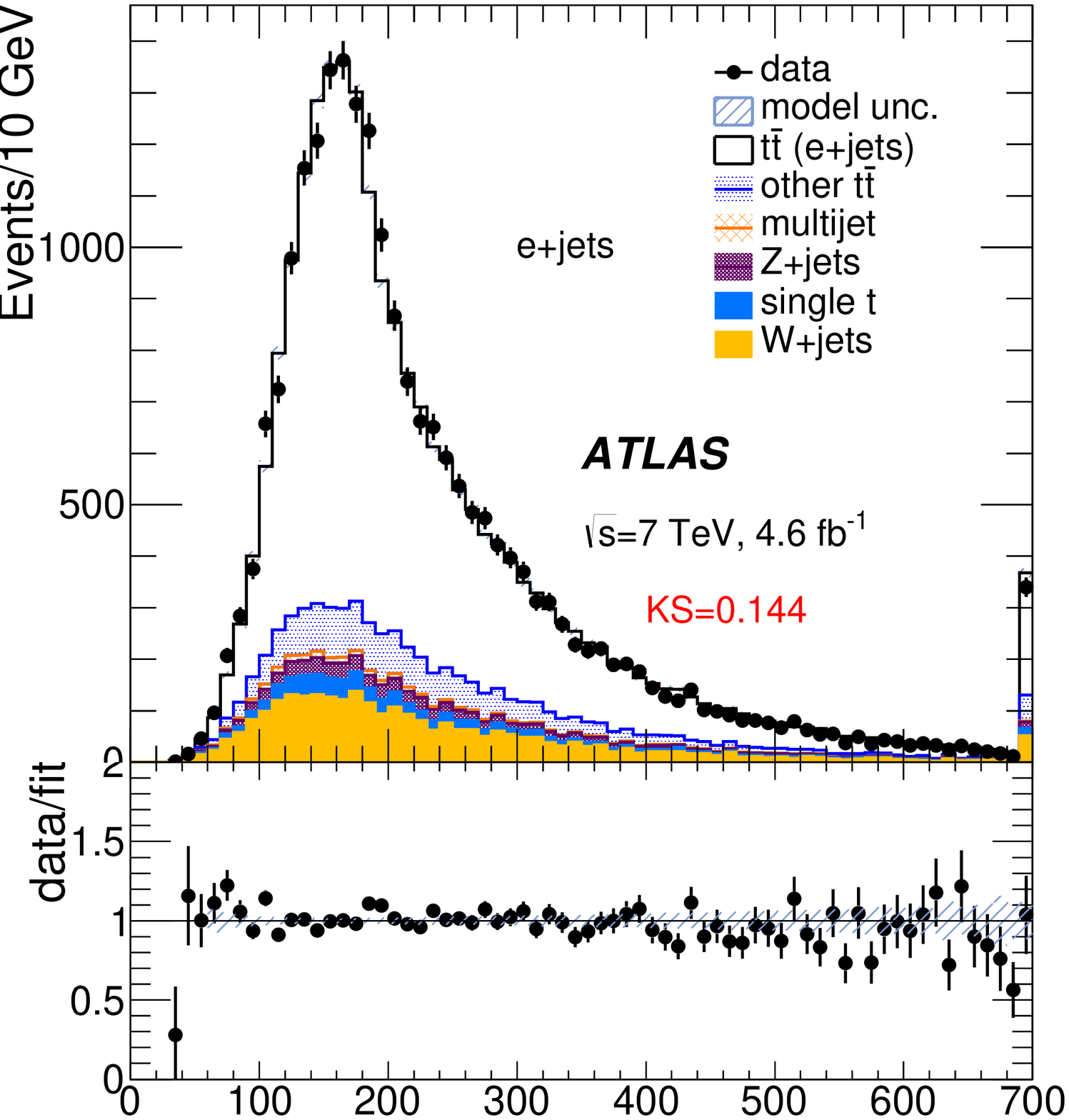,width=7.5cm, height=6.5cm} & 
\epsfig{figure=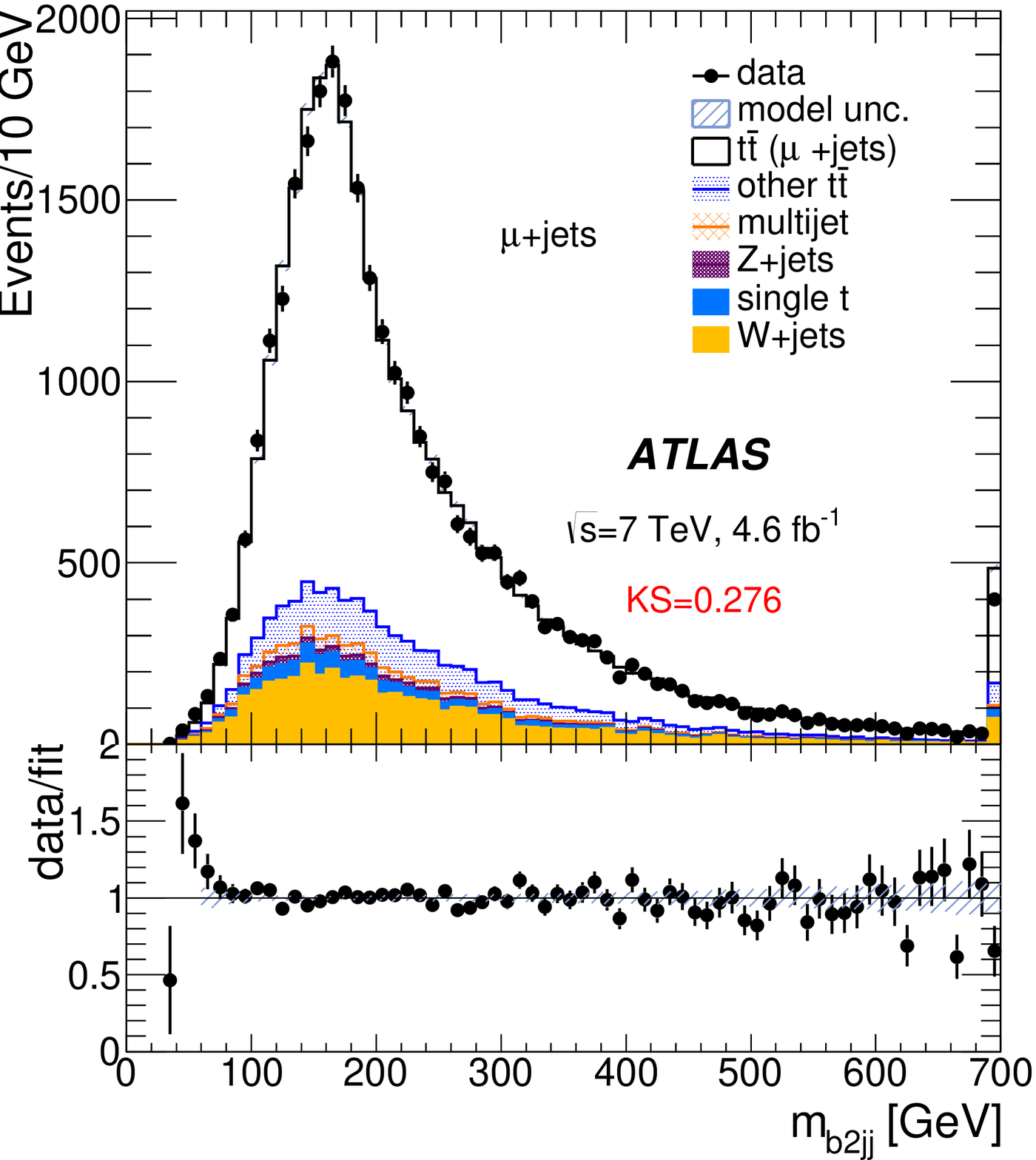,width=7.5cm, height=6.5cm} \\ 
(e) $e$+jets& (f) $\mu$+jets\\
\end{tabular}
\caption{\label{fig:ljets_fit1} Distributions in data compared to the SM expectations after fitting the following distributions: (a,b) the invariant mass of two highest-$\pt$ jets not designated as $b$-jets; (c,d) the invariant mass of the leading jet designated as $b$-jet and the jets used for $m_{jj}$ ($m_{b1jj}$), and (e,f) the invariant mass of the second jet designated as a $b$-jet and the two jets used for $m_{jj}$ ($m_{b2jj}$). The distributions are shown for events with isolated leptons, at least four jets, at least one $b$-tag, and $\met>30$ GeV, with the $e$+jets and $\mu$+jets channels separated. The last bin shows the overflow. The ratio plots show the result of dividing the data points by the model expectation. The model uncertainty (model unc.) is the sum in quadrature of the statistical uncertainties of the templates used in the fits. KS is the value of the Kolmogorov-Smirnov goodness-of-fit test.}
\end{figure}

\begin{figure}[!hbt]
\begin{tabular}{cc}
\epsfig{figure=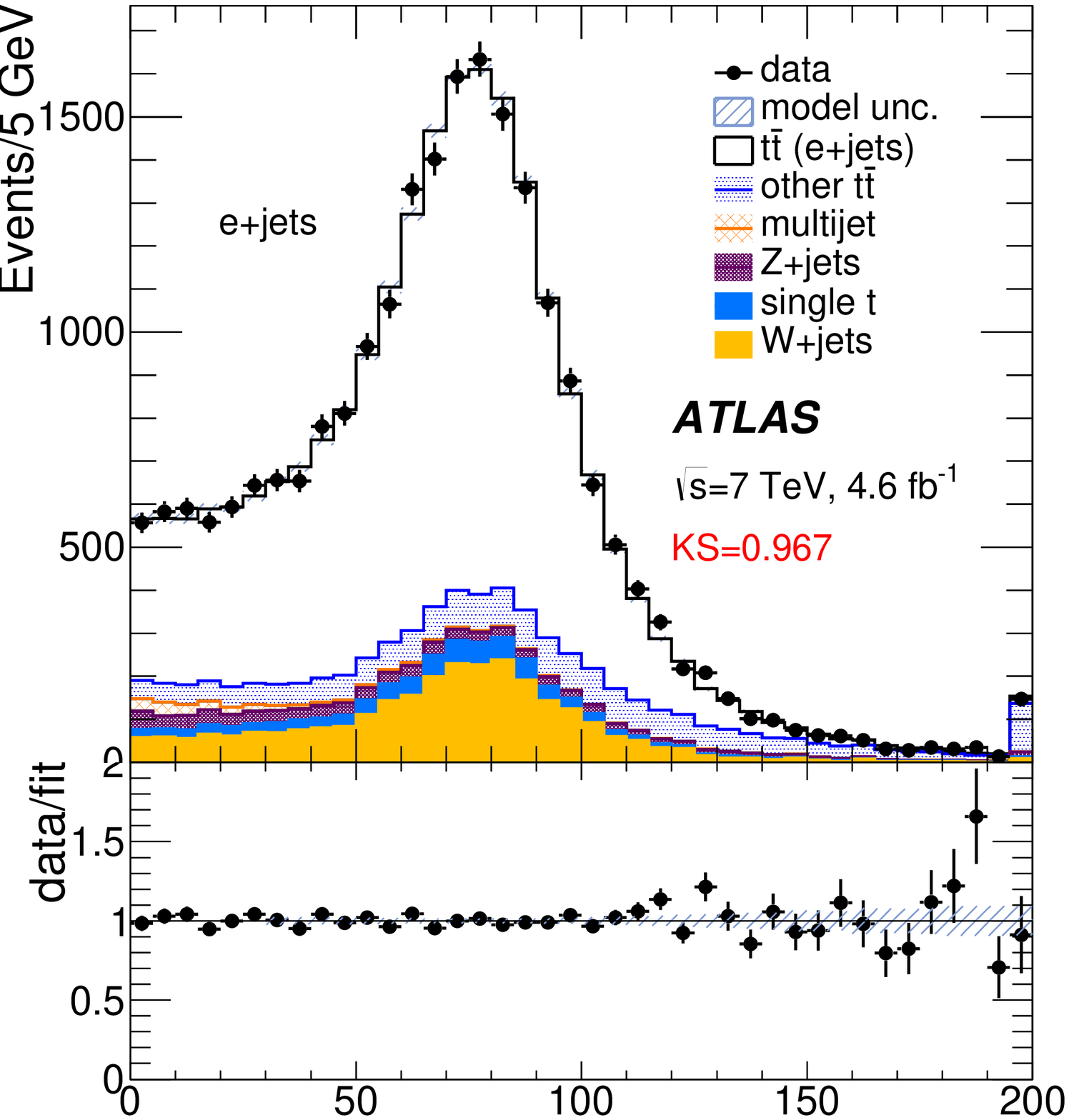,width=7.5cm} & 
\epsfig{figure=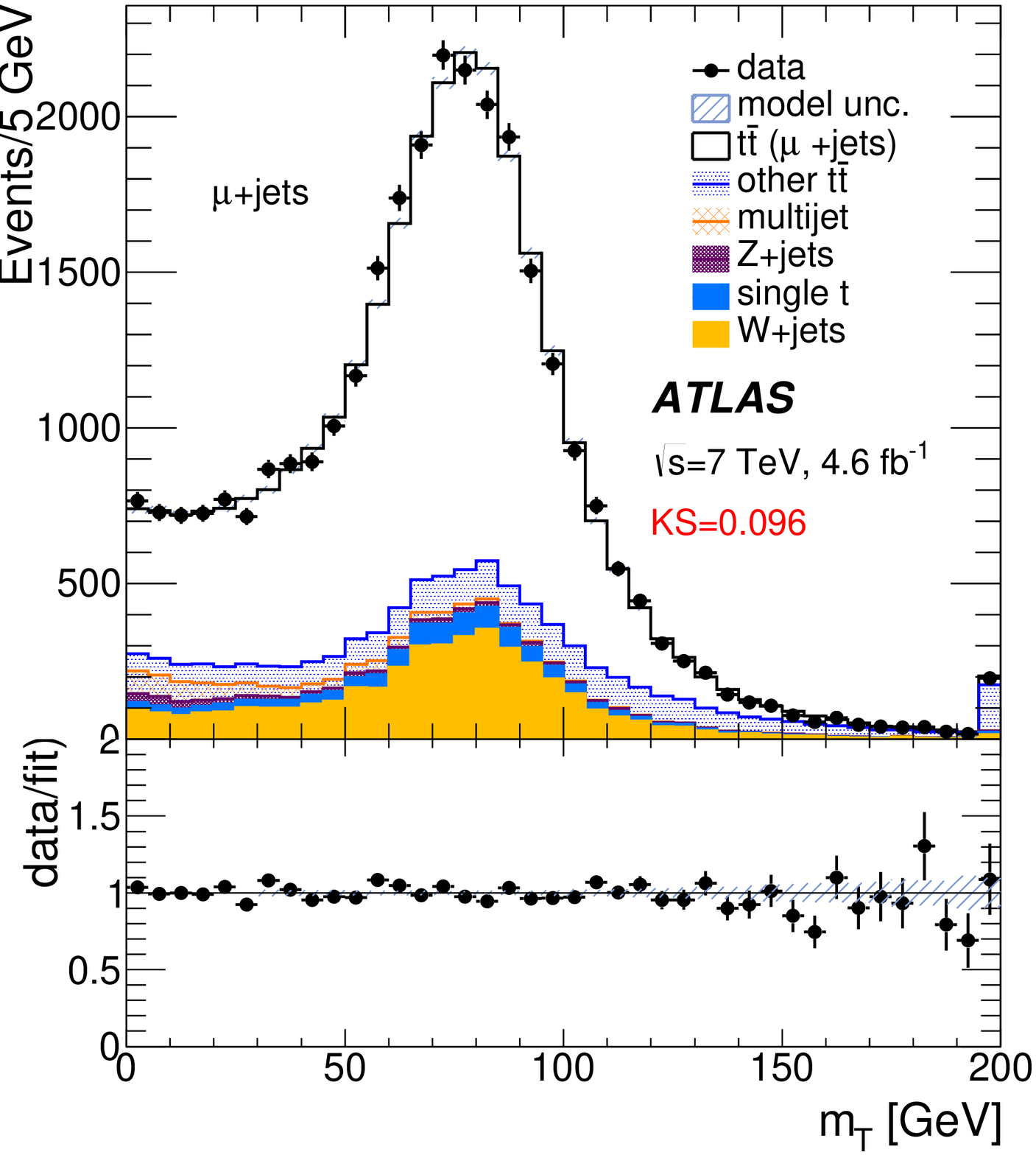,width=7.5cm} \\
\end{tabular}
\caption{\label{fig:ljets_fit2}  The transverse mass of lepton and $\met$ ($m_{\rm T}$) distributions 
for events with isolated leptons, at least four jets, at least one $b$-tag  and $\met>30\gev$ in
the $e$+jets and $\mu$+jets channels. The last bin shows the overflow. The ratio plots show 
the result of dividing the data points by the model expectation. The model uncertainty (model unc.) is the sum in quadrature of the statistical uncertainties of the templates used in the fits. KS is the value of the Kolmogorov-Smirnov goodness-of-fit test.}
\end{figure}

   The $m_{\rm T}$ distributions for events with $\met<30$ GeV, used in the fits, 
are shown in Fig.~\ref{fig:ljets_Mt}.
 Table~\ref{tab:ljet_fits} shows the predicted contributions from each channel, 
combining events with $\met<30\gev$ and $\met>30\gev$.
Figure~\ref{fig:ljets_fit1} 
 shows that the fits describe well the full $e$+jets and $\mu$+jets 
event distributions 
of $m_{jj}$, $m_{b1jj}$ and
$m_{b2jj}$ after requiring
$\met>30\gev$. Figure~\ref{fig:ljets_fit2} shows  
 the $m_{\rm T}$
distribution for events with $\met>30\gev$ compared with the predicted contributions, which agree well with the data.

Noticeable features from these fits are as follows:
\begin{itemize}
\item The largest backgrounds originate in $W$+jets (15\%) and other $t \overline{t}$ channels
(8.5\%); the rest add up to 12\% (multijets 5.3\%, $Z$+jets 3.9\%, and single top 3.0\%).
\item The numbers of $t \overline{t}$ and $W$+jets events obtained by fitting are 
in good agreement with those predicted by the SM.
\end{itemize}

\begin{table}[!hbt]
\centering
\caption{\label{tab:ljet_fits} Results from fitting $e$+jets and $\mu$+jets mass distributions 
from $\ell$+jets events requiring exactly one isolated lepton ($e$ or $\mu$),  at least four jets, and at least one $b$-tag.
The uncertainties quoted here are from the fits and do not include systematic uncertainties. The $Z$+jets 
contribution is normalized  to the MC expectation. In addition to MC statistical uncertainty, it includes 
the uncertainty from the scale factors applied to the simulation in order to match the jet multiplicity and the Z boson $\pt$ dependence to that observed in the data . The single top and diboson contributions are normalized 
to MC predictions, include only MC statistical uncertainty and the SM cross section uncertainty. The (MC) rows give the numbers expected from MC simulation. The $\chi^2/ndf$ row gives the $\chi^2$ and degrees of freedom of the fits.}

\begin{tabular}{l|cc}
\multicolumn{3}{c}{} \\ 
\hline
\hline 
Channel & $e$+jets & $\mu$+jets \\
\hline
$t \overline{t}\rightarrow\ell$+jets & 19710$\pm$280 & 25090$\pm$310 \\
(MC)                         & (18966$\pm$\phantom{0}31) & (24233$\pm$\phantom{0}34) \\
\hline
\hline
$t \overline{t}$ (other)      & \phantom{0}2674$\pm$\phantom{0}30  & \phantom{0}3393$\pm$\phantom{0}30\\
(MC)                         & (\phantom{0}2577$\pm$\phantom{0}11) & (\phantom{0}3277$\pm$\phantom{0}16) \\ 
\hline
$W$+jets                     &  \phantom{0}4800$\pm$500 &  \phantom{0}5600$\pm$500 \\ 
(MC)                           & (\phantom{0}4140$\pm$\phantom{0}70)& (\phantom{0}5850$\pm$\phantom{0}90) \\  
\hline
$Z$+jets (MC)                & \phantom{0}1900$\pm$500  &  \phantom{00}790$\pm$200 \\
Single top (MC)              & \phantom{00}910$\pm$\phantom{0}70   & \phantom{0}1170$\pm$\phantom{0}80 \\
Diboson (MC)                 & \phantom{000}5.0$\pm$\phantom{0}0.2   & \phantom{000}6.1$\pm$\phantom{0}0.2  \\
Multijet                    & \phantom{0}1000$\pm$120  & \phantom{0}2800$\pm$140 \\
\hline
\hline
Total background                   & 11333$\pm$700 & 13700$\pm$600 \\
Signal+background                  & 31000$\pm$800 & 38800$\pm$700 \\
\hline
Data                         & 30733 & 40414 \\
\hline
\rule{0pt}{12pt}$\chi^2/ndf$ & 188/207 & 218/207\\
\hline
\hline
\end{tabular}

\end{table}

\clearpage

\section{Dilepton + jets channel}
\label{sec:l+l}
\begin{figure}[!hbt]
\begin{tabular}{cc}
\epsfig{figure=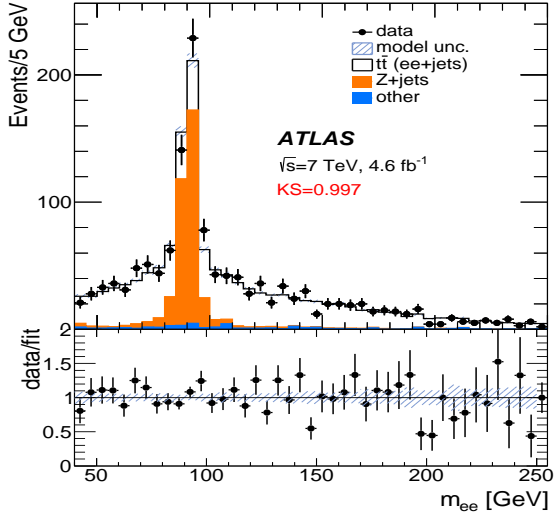,width=7.5cm, height=7.0cm} & 
\epsfig{figure=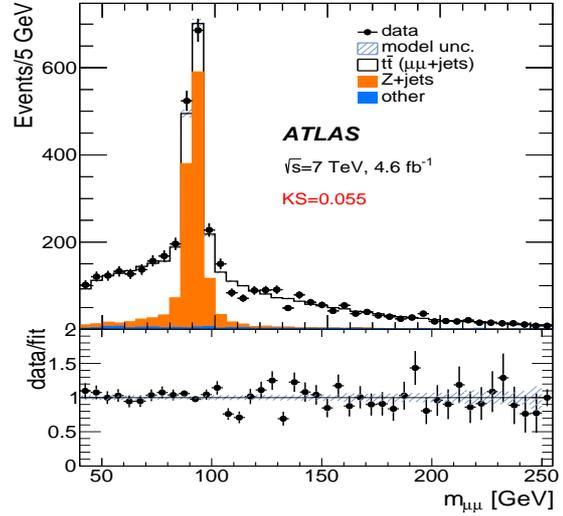,width=7.5cm, height=7.0cm} \\
(a) $ee$+jets & (b) $\mu\mu$+jets \\
\epsfig{figure=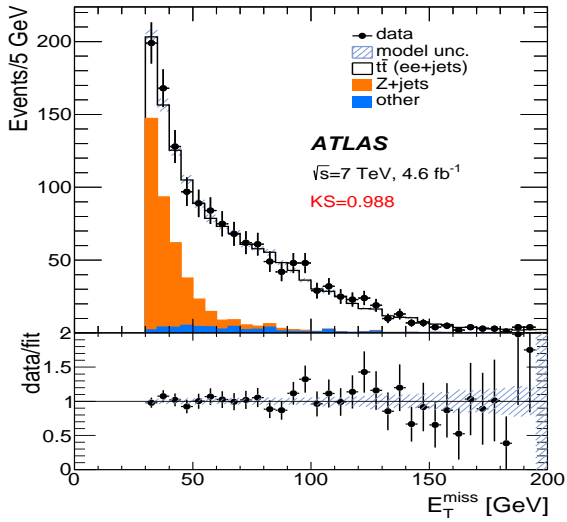,width=7.5cm, height=7.0cm} & 
\epsfig{figure=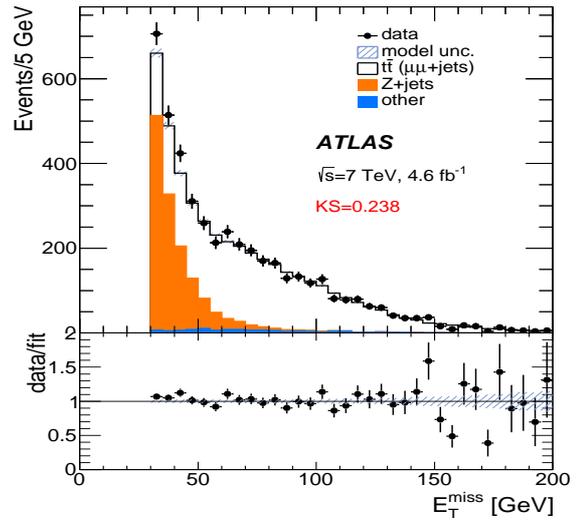,width=7.5cm, height=7.0cm} \\
(c) $ee$+jets & (d) $\mu\mu$+jets \\
\end{tabular}
\caption{\label{fig:dil_fit} Dilepton invariant masses (a) $m_{ee}$, (b) $m_{\mu\mu}$, and $\met$ distributions for events with two isolated
leptons, $\met>30\gev$, at least two jets, and at least one $b$-tag in the (c) $ee$+jets and (d) $\mu\mu$+jets 
channels.
The $Z$+jet entries 
include a small contribution from $Z\rightarrow\tau^+\tau^-$ with both $\tau$ leptons
decaying to $e$ or $\mu$. The ratio plots show the result of dividing the data points by the fit. The model uncertainty (model unc.) is the sum in quadrature of the statistical uncertainties of the templates used in the fits. KS is the value of the Kolmogorov-Smirnov goodness-of-fit test.}
\end{figure}

\begin{figure}[!hbt]
\begin{tabular}{cc}
\epsfig{figure=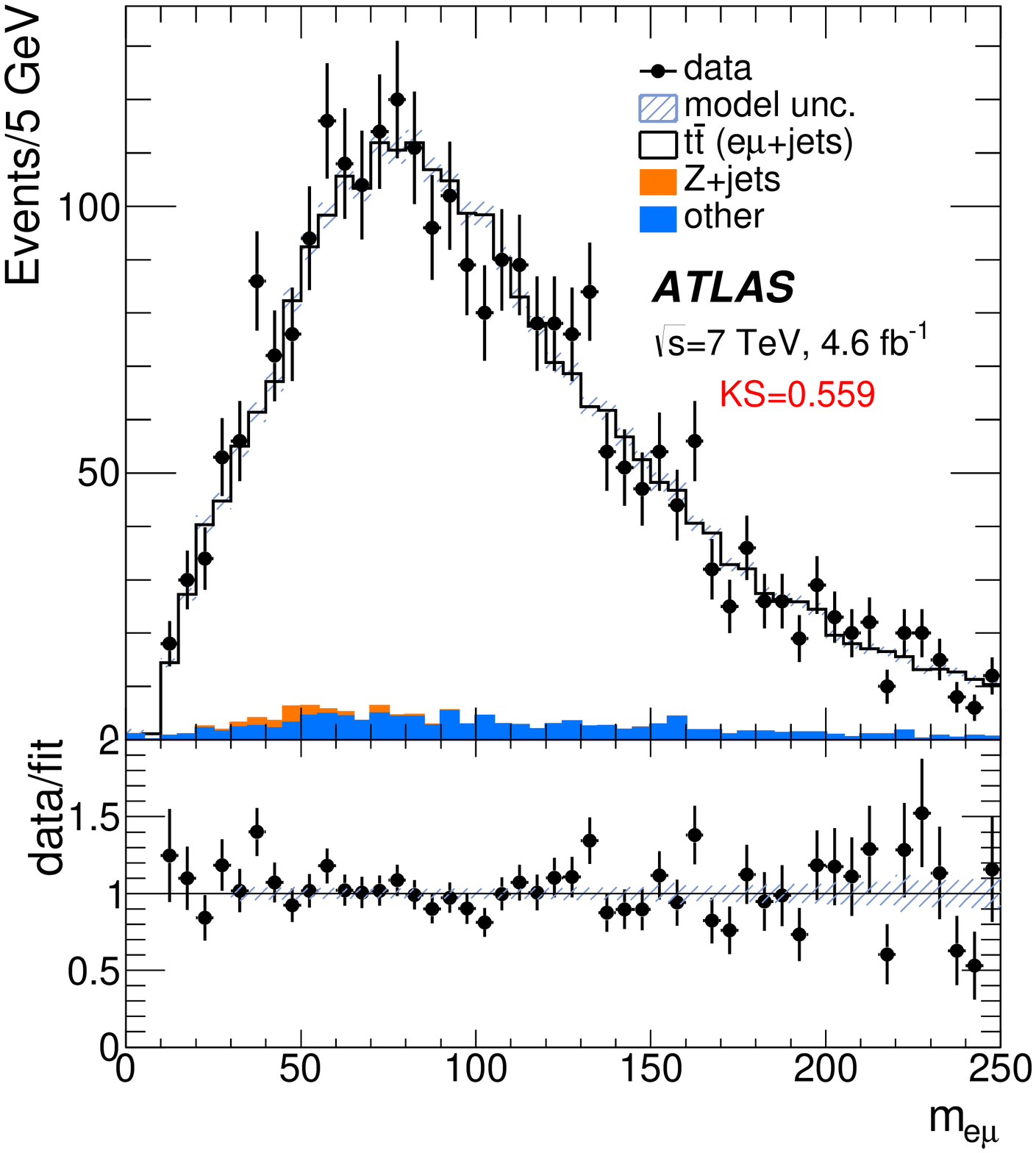,width=7.5cm, height=7.0cm} & 
\epsfig{figure=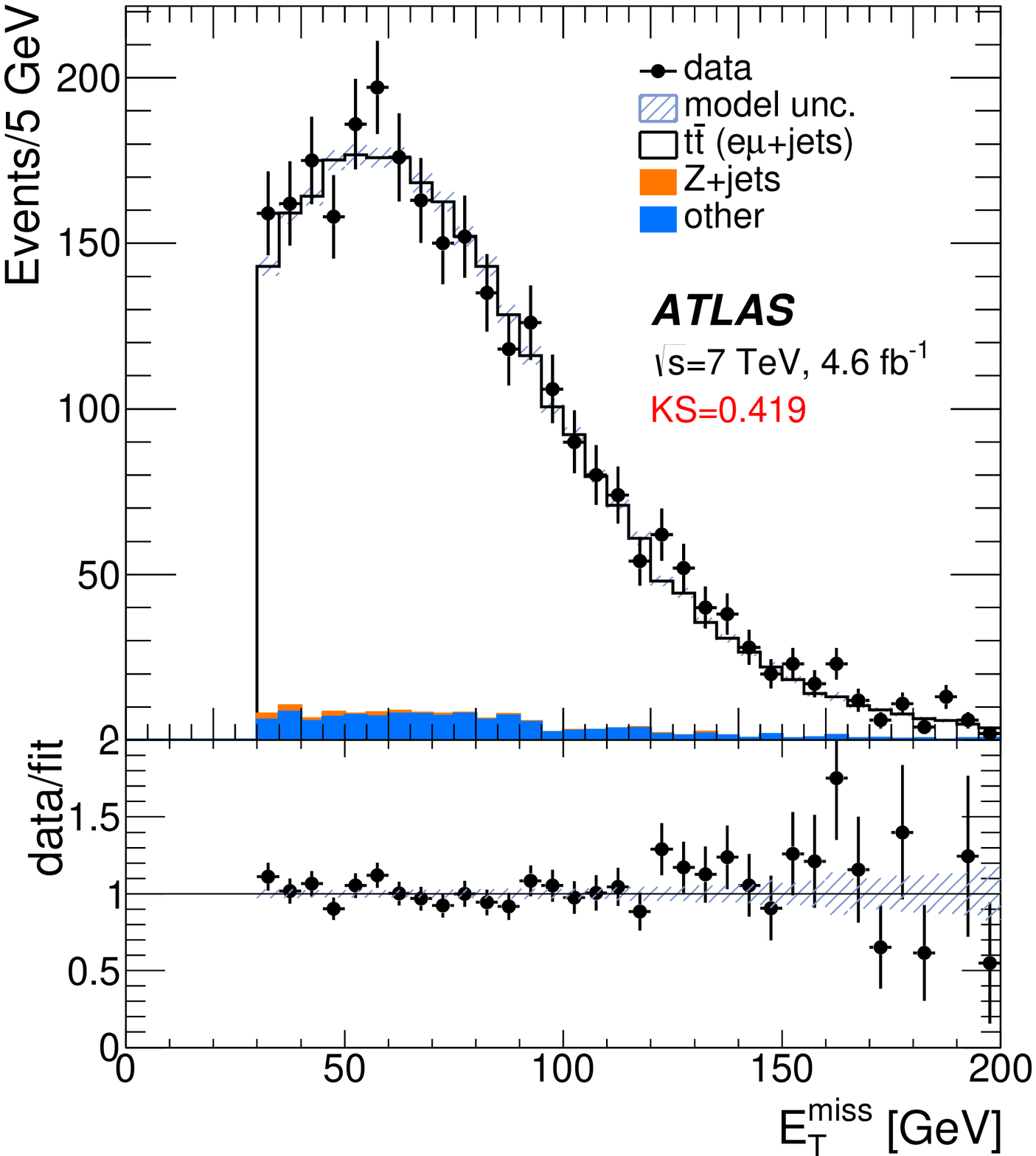,width=7.5cm, height=7.0cm} \\
(a)  & (b) \\ 
\end{tabular}
\caption{\label{fig:emu_fit} (a) Invariant mass of electron and muon ($m_{e\mu}$)  and (b) $\met$ distributions for 
$e\mu$ events after requiring one isolated $e$ and one isolated $\mu$, 
$\met>30\gev$, at least two jets, and at least one $b$-tag. The ratio plots show the result of dividing the data points by the fit. The model uncertainty (model unc.) is the sum in quadrature of the statistical uncertainties of the templates used in the fits. KS is the value of the Kolmogorov-Smirnov goodness-of-fit test.}
\end{figure}

The number of $t \overline{t}\rightarrow \ell\ell'$+jets events in the data is extracted 
by fitting two  dilepton invariant mass distributions: one with $30<\met<60$ GeV
and the other with $\met>60\gev$. 
The most significant background to the $t \overline{t}\rightarrow \ell\ell'$+jets 
channels after requiring $\met>30\gev$ and at least one $b$-tagged jet comes 
from the $Z (\rightarrow \ell \ell')$+jets, with a smaller contribution from single top production (4\%).
Since the $\met$ distribution falls more rapidly for the $Z$+jets background than for the $t \overline{t}$
 signal process separating it into two $\met$ bins improves the sensitivity of the 
fit for separating the two processes.
Backgrounds from dibosons and jets misidentified as isolated leptons (mainly
from W+jets with leptons from heavy-quark semileptonic decays or an isolated
charged hadron misidentified as a lepton, together denoted as nonprompt leptons) amount to 1.0\% of the events.  
The background
from nonprompt isolated leptons is estimated from the number of data events 
with lepton pairs with the same charge after subtracting a very small 
expected contribution from diboson processes.
The invariant mass distributions are fitted with three templates: 
one derived from a $t \overline{t}$ MC sample, one from a $Z$+jets MC sample, and one summed 
over all other
contributions. Only the amounts contributed by $t \overline{t}$ and $Z$+jets
are allowed to vary. 
The $Z$ boson background in the $e\mu$+jets channel from the $Z(\rightarrow \tau\tau\rightarrow e\mu)$+X
channel is too small to be 
extracted by a fit, so $m_{e\mu}$ is fitted only 
for the number of $t \overline{t}$ events in the data while the background is fixed.
The fits in the $\ell\ell$ channel are performed over a mass range from 
$40\gev$ to $250\gev$ and in the $e\mu$ channel over a mass range from $10\gev$ to $250\gev$. 
Figures~\ref{fig:dil_fit} and \ref{fig:emu_fit} show that the $m_{\ell\ell'}$ and $\met$ distributions are
well described in all dilepton channels. Results of the fits are given in Table~\ref{tab:dil_fits}.

\begin{table}[!hp]
\centering
\caption{\label{tab:dil_fits} Results from fitting $\ell\ell'$ invariant mass distributions using two $\met$ regions from $\ell\ell'$+jets events requiring two isolated leptons ($e$ or $\mu$), $\met>30$ GeV, at least two jets, and at least one $b$-tag. The numbers of events are after summing events from both $\met$ regions  
$\met<60$ GeV and $\met>60$ GeV.
The uncertainties are from the fits and do not include systematic uncertainties. 
 The single top and diboson contributions are normalized to the SM predictions and include only the MC statistical
uncertainty and the uncertainty on the SM cross section. The (MC) rows give the numbers expected from MC simulation.}
\begin{tabular}{l|ccc}
\multicolumn{4}{c}{} \\ 
\hline
\hline 
Channel & $\mu\mu$+jets & $ee$+jets & $e\mu$+jets \\
\hline
$t \overline{t}$ & 2890$\pm$\phantom{0}80  & 1000$\pm$40 &  2640$\pm$50\\
(MC)             & (2536$\pm$\phantom{0}11)& (\phantom{0}903$\pm$\phantom{0}6) & (2420$\pm$11) \\
\hline
$Z$+jets         & 1380$\pm$\phantom{0}50 & \phantom{0}379$\pm$11 & \phantom{00}13$\pm$\phantom{0}4 \\
(MC)             & (1267$\pm$\phantom{00}8)& (\phantom{0}385$\pm$11)& (\phantom{00}13$\pm$\phantom{0}4) \\
\hline
Single top (MC)  & \phantom{00}86$\pm$\phantom{00}8   & \phantom{00}36$\pm$\phantom{0}7& \phantom{00}98$\pm$\phantom{0}9\\
Diboson (MC)  & \phantom{00}22$\pm$\phantom{00}1   & \phantom{00}8.1$\pm$0.5 & \phantom{00}3.3$\pm$0.3\\
Fake leptons     &  \phantom{00}17$\pm$\phantom{0}10 & \phantom{00}17$\pm$\phantom{0}8 & \phantom{00}19$\pm$10\\
\hline
\hline
Total background       & 1430$\pm$\phantom{0}50 & \phantom{0}442$\pm$15 & \phantom{0}136$\pm$12 \\
                 
Signal+background     & 4400$\pm$100 & 1440$\pm$40 & 2770$\pm$80 \\
                 
\hline
Data             & 4102 & 1447 & 2848 \\
\hline
\rule{0pt}{12pt}$\chi^2/ndf$ & 35/34 & 31/34 &58/49 \\
\hline
\hline
\end{tabular}

\end{table}

\clearpage

\section{Lepton+$\tau_\mathrm{had}$+jets channel}
\label{sec:l+tau}
Unlike the single-lepton + jets and dilepton channels 
the background in the $\ell\tau_\mathrm{had}$+jets channel
is not small and is dominated by contributions from other $t \overline{t}$ channels.
Thus, invariant masses and other kinematic variables are not sufficiently sensitive
to separate signal and background. In this case 
a BDT multivariate discriminant, named BDT$_j$, is used to separate $\tau$ leptons from jets identified
as $\tau$ candidates (see
Sec.~\ref{sec:common_objects}). Compared to the previous ATLAS measurement 
 with this channel~\cite{PLB}, the present analysis uses only one-prong $\tau$ decays and  
is based on a larger data sample with a different 
background model to reduce the statistical uncertainty on the background prediction. 

\subsection{Tau background templates}
\label{sec:taubackground}
In order to separate the contribution of processes with $\tau$ leptons (signal) from 
those with jets misidentified as $\tau$ (fake $\tau$) the BDT$_j$ distributions 
of selected events are fitted with templates for fake $\tau$ distributions derived from data
and true $\tau$ lepton distributions derived from MC simulation. Control data samples 
to obtain templates of jets misidentified as 
$\tau$ candidates
are selected with the following requirements:
\begin{itemize}
\item exactly one isolated electron with $\pt^e>25\gev$ and no identified muons for the $e+\tau$ channel; 
\item or exactly one isolated muon with $\pt^{\mu}>20\gev$ and no identified electrons for the $\mu+\tau$ channel; 
\item and no additional muons with $\pt>4\gev$; 
\item and $40\gev <m_{\rm T}(\ell,\met)<100\gev$;
\item and exactly one $\tau$ candidate and at most one additional jet.
\end{itemize}

 There are two mutually exclusive control samples:
\newline 
The $W$+1-jet sample contains a lepton, one jet misidentified as a $\tau$ candidate and no additional jets.
The $W$+2-jets sample contains a lepton and exactly two jets with the lower $\pt$ jet misidentified as a $\tau$ candidate.

The control samples are divided into two subsamples, one with $\tau$ and $\ell$
having the opposite-sign charges (OS), and the other with $\tau$
and $\ell$ having the same-sign charges (SS). 
The $W+1$-jet sample is rich in jets originating from quark hadronization (quark jets) while the $W+2$-jets
sample has a high percentage of jets originating from gluon hadronization 
(gluon jets) as determined from MC studies. 
One can extract the distributions
of gluon jets misidentified as $\tau$ candidates since the number
of gluon jets in OS and SS samples must be the same because they are not correlated with the charge of the lepton.
Fake $\tau$ template shapes depend on the jet type. Those from 
light-quark jets
peak at higher BDT$_j$ values than those from gluon jets. The signal contributes only
to OS events. Therefore,  
the BDT$_j$ distributions of OS events are fitted with a pair of background 
templates, whose linear combination equals the sum of the OS light-quark and 
gluon jets identified as $\tau$ candidates, and a signal $\tau$ template.
MC studies show that requiring $\tau$ candidates that have only one 
associated charged particle strongly suppresses jets originating from
heavy quarks ($c$-jets, $b$-jets). The $b$-jets are further suppressed by excluding $\tau$ candidates that are tagged as $b$-jets. The BDT$_j$ template from remaining $c$-jets identified as $\tau$ candidates is similar to
the light-quark template.
 The signal template is constructed
 by summing the expected contribution of any channel that has a real $\tau$
lepton or a lepton misidentified as a $\tau$ lepton.

\begin{figure}%[!hbt]
\begin{tabular}{ccc}
\epsfig{figure=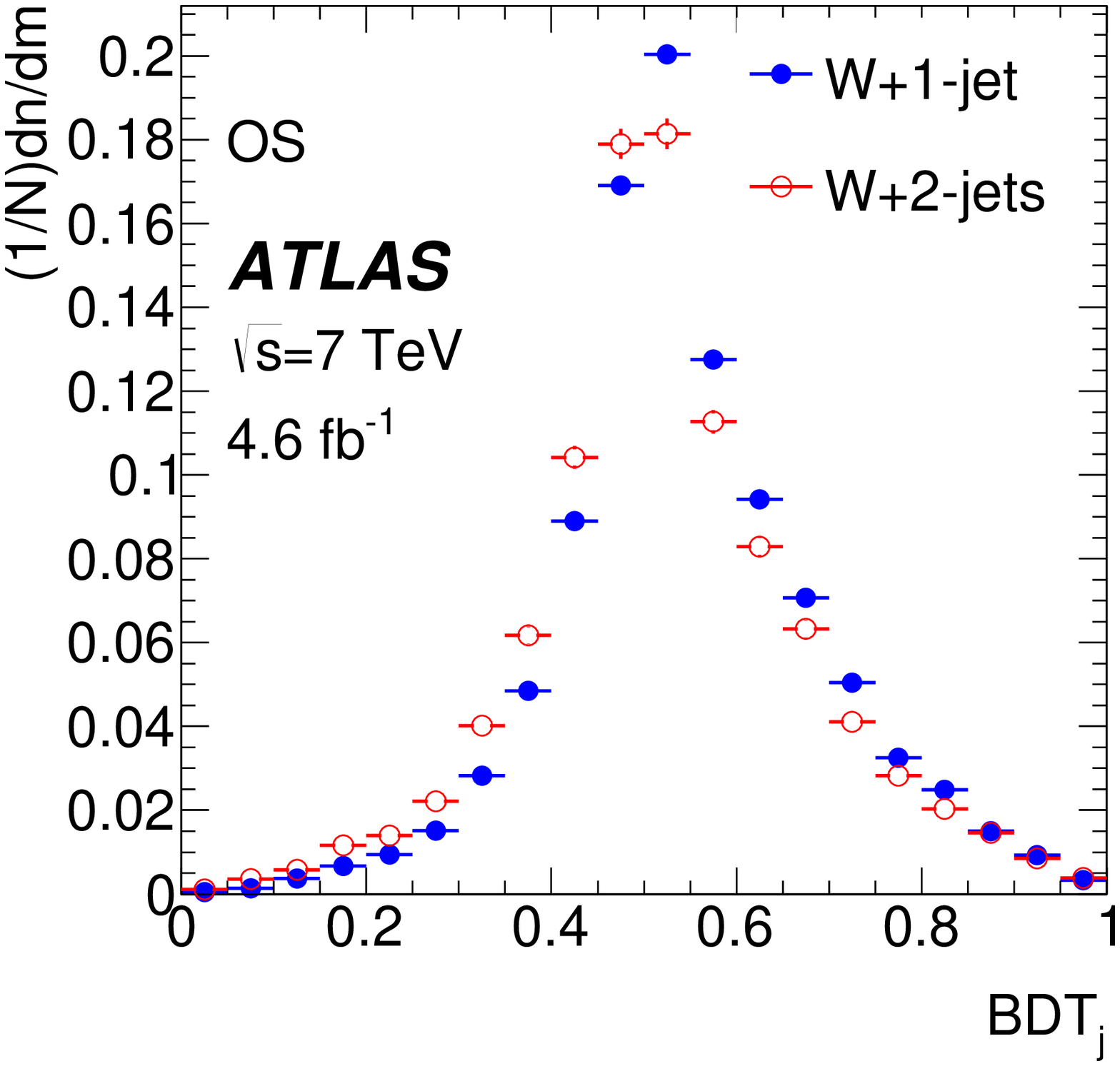,width=5.0cm}  & \epsfig{figure=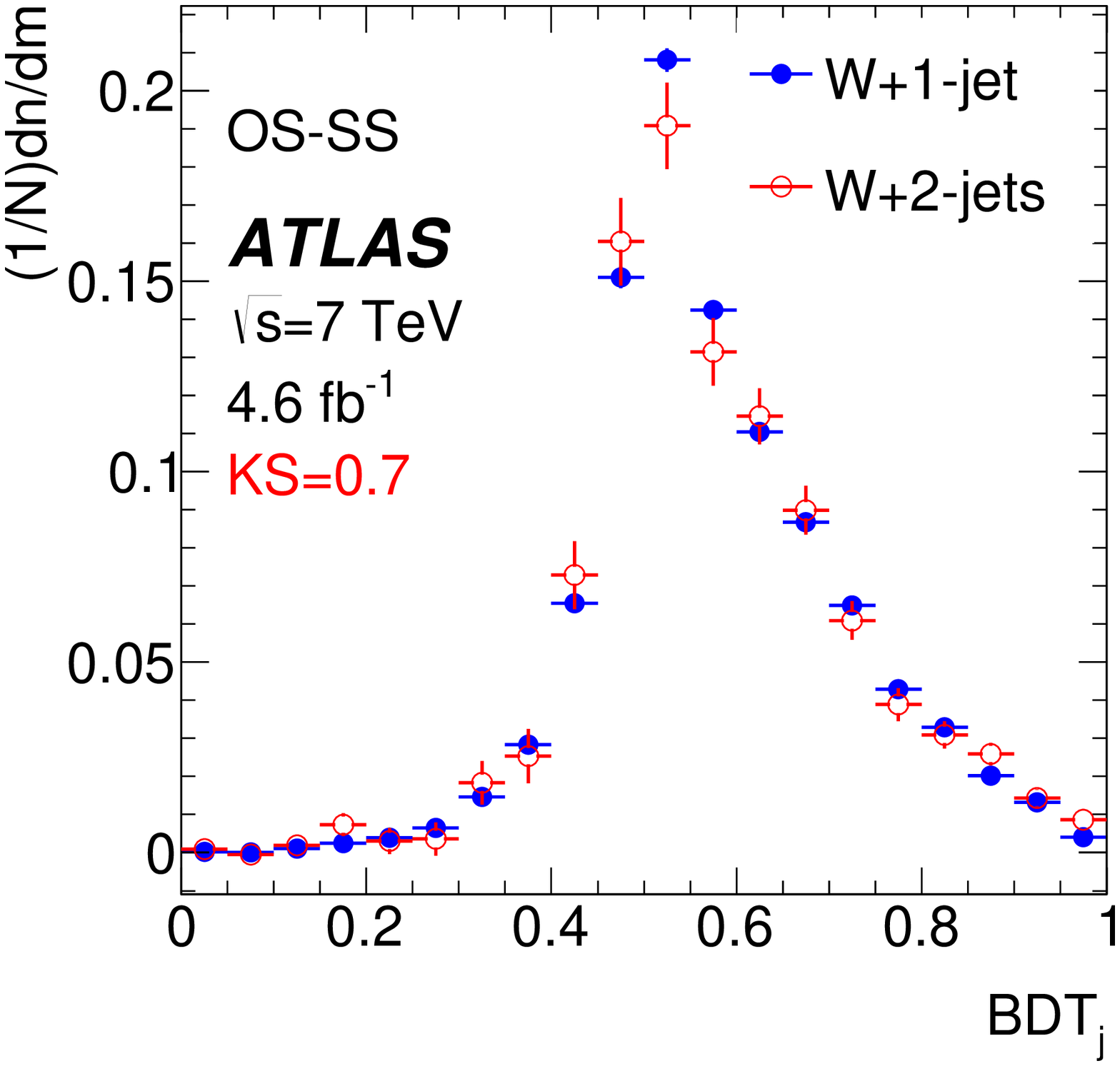,width=5.0cm}
&  \epsfig{figure=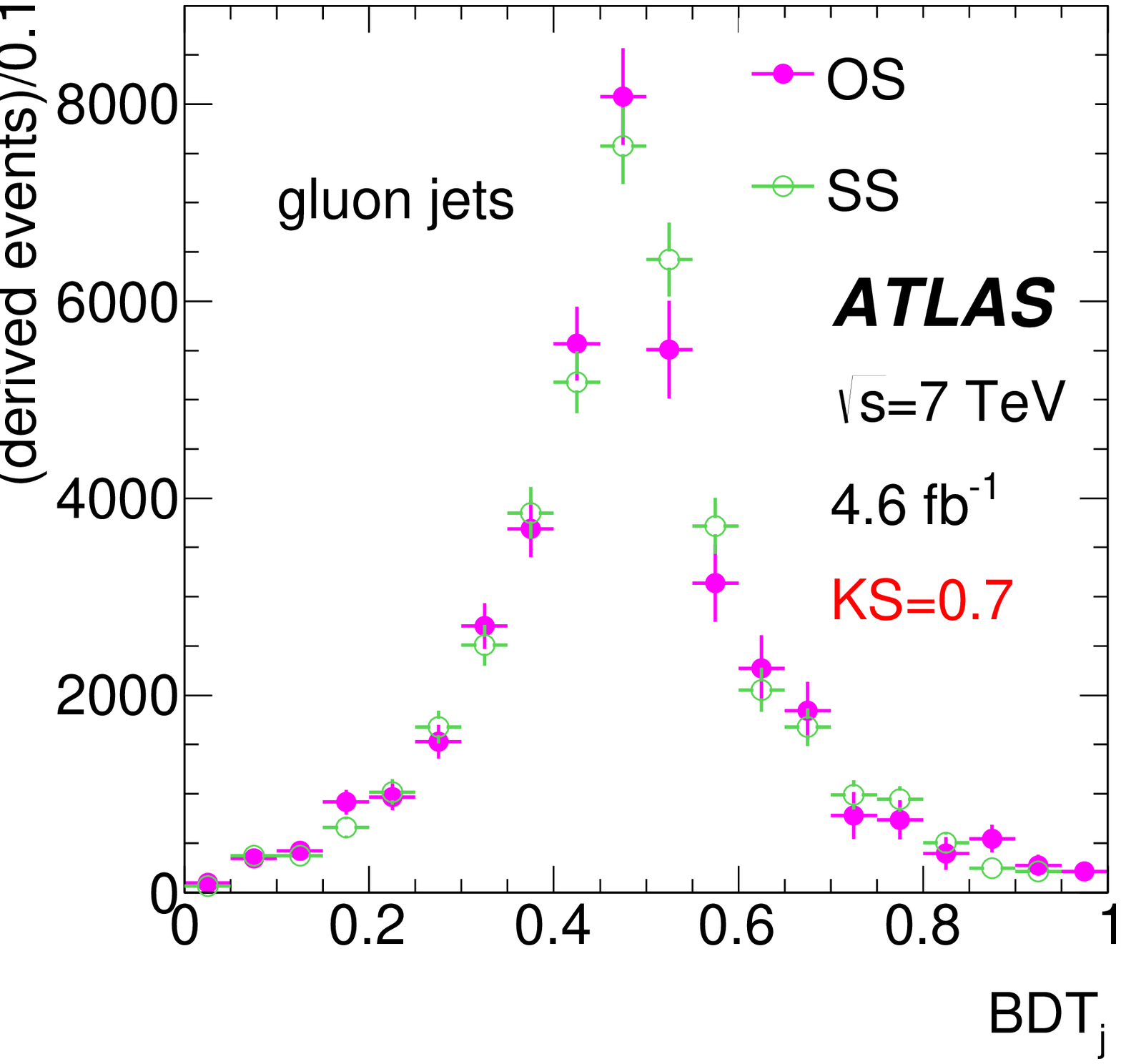,width=5.0cm}  \\
\end{tabular}
\caption{\label{fig:BDT12} Normalized distributions of the output of the boosted decision tree used to discriminate $\tau$ leptons from jets misidentified as $\tau$s, BDT$_j$, for $\tau$ candidates from $W+1$-jet and $W+2$-jets samples for 
leptons with opposite sign (OS), the distribution of opposite-sign leptons with the same-sign lepton distribution subtracted (OS$-$SS), and the extracted BDT$_j$ distributions ($K\cdot G$, see text)  for gluon jets misidentified as $\tau$ candidates is shown.}
\end{figure}

In the $W+2$-jets sample
the lower-$\pt$ jet has a high probability of coming from final- or initial-state radiation
and thus a high probability of being a gluon jet. 
In the following, OS1 (SS1) stands for the $\tau$ fake BDT$_j$ distribution obtained from OS (SS) $W+1$-jet 
data sets and OS2 (SS2) represent the equivalent distribution for $W+2$-jets. 
Figures~\ref{fig:BDT12}(a) and \ref{fig:BDT12}(b) show the OS and OS-SS distributions normalized to compare the shapes. 
It can be seen that there are significant differences between OS1  
and OS2, 
but if one subtracts the SS distribution from the OS distribution (OS-SS) 
the  shapes are in good agreement. The distributions are a sum 
of light-quark  jets and gluon jets, and can be described by the following equations:
\begin{equation}\label{eq:os1}
{\rm OS1}=a_1 \cdot {\rm OS_q}+b_1 \cdot G,
\end{equation}
\begin{equation}\label{eq:ss1}
{\rm SS1}=c_1 \cdot {\rm SS_q}+b_1 \cdot G,
\end{equation}
\begin{equation}\label{eq:os2}
{\rm OS2}=a_2 \cdot {\rm OS_q}+b_2 \cdot G,
\end{equation}
\begin{equation}\label{eq:ss2}
{\rm SS2}=c_2 \cdot {\rm SS_q}+b_2 \cdot G,
\end{equation}
where $\rm OS_q$ ($\rm SS_q$) is a function describing the shape of the distribution
of light-quark jets contributing to OS (SS) and $G$ is the corresponding
function for gluon jets.
The observation that the OS1$-$SS1 and OS2$-$SS2 distributions have the same shape 
leads to the conclusion that
$a_1/c_1=a_2/c_2$ for any $E_{\rm T}$ as the $E_{\rm T}$ of $\tau$ candidates from $W+2$-jets 
are significantly lower than those from $W+1$-jet. Using the above equations, one can extract the $G$ function from the OS and SS distributions separately, i.e.
\begin{equation}\label{eq:gl1}
K \cdot G=(R \cdot {\rm OS2} - {\rm OS1}),
\end{equation}
\begin{equation}\label{eq:gl2}
K \cdot G=(R \cdot {\rm SS2} - {\rm SS1}),
\end{equation}
where $R$ is the ratio of the total number of OS1$-$SS1 events to OS2$-$SS2 events
and $K=R\cdot b_2 -b_1$ is an unknown constant that must be the same whether
SS or OS is used to extract $G$. Figure \ref{fig:BDT12}(c) shows the extracted
$K\cdot G$ distributions for $\tau$ candidates. It is seen that the OS and SS distributions
are fully consistent with each other and can be summed to reduce the statistical uncertainties.

In principle any background BDT$_j$ distribution can be described by a linear combination of $G$ and OS1 distributions.  Furthermore, the BDT$_j$ distributions
depend on $E_{\rm T}$ of the $\tau$ candidates, which differs from sample to sample. 
The $E_{\rm T}$ dependence of the BDT$_j$ is taken 
into account by fitting separate $E_{\rm T}$ regions with templates derived for 
those regions weighted to reproduce the $E_{\rm T}$ distributions of 
the expected background. The OS1 sample has a small (2\%) number of $\tau$ leptons from dibosons and $Z\rightarrow \tau^+\tau^-$ final states that 
have no impact on the fits to $\ell\tau_\mathrm{had}$+jets BDT$_j$ data distributions 
whether or not they are subtracted from the OS1 template.

\newpage

\subsection{Signal extraction by fitting to BDT$_j$ shape}
\label{sec:fits}

The final background normalization and signal measurement are
established through fitting templates to the data.  
There are various classes of background: 
\begin{enumerate}
\item from processes with an isolated $\ell$ where a jet is misidentified
as a $\tau$ candidate;
\item from processes other than $t \bar t$
that have $\tau$ leptons and an isolated $\ell$;
\item from processes with two isolated $\ell$ where one $\ell$ is misidentified
as a $\tau$ candidate;
\item from multijet processes where both $\ell$ and $\tau$  are from one jet
misidentified as an isolated $\ell$ and another as a $\tau$ candidate.
\end{enumerate}

The dominant background
to the $t \overline{t}\rightarrow\ell \tau_\mathrm{had}$+jets channel comes from the 
$t \overline{t}\rightarrow\ell$+jets channel 
with one jet misidentified as a $\tau$ candidate (class 1). The only powerful
suppression technique for that background is $\tau$ identification, thus the best variable is the BDT$_j$
score, described with the $\tau$ candidate selection in 
Sec.~\ref{sec:common_objects}. Background of classes 1 and 4 is taken into account using templates  
consisting of light-quark jet $\tau$ fakes
and gluon jet $\tau$ fakes derived from enriched $W$+jets data samples as
described in Sec.~\ref{sec:taubackground}.

The signal BDT$_j$ template is derived from MC $\tau$ candidates 
that are matched to a $\tau$ lepton or a lepton from MC events
that satisfy the event selection (classes 2 and 3) .
The class 2 processes contributing to the signal template are: $t\bar{t}\rightarrow \ell\tau$+jets, $Z(\rightarrow\tau^+\tau^-)$ +jets,
and small contributions from single top and diboson events.
The main backgrounds of class 3 are $Z\rightarrow e^+e^-$ and $t \bar t$ events.
Most electrons are removed by the BDT$_e$ cut (see Sec.~\ref{sec:common_objects}); 
the few that remain are indistinguishable
from $\tau$ leptons. There is an even smaller number of muons 
overlapping with $\tau$ candidates that are not removed by the muon veto and are
also indistinguishable from $\tau$ leptons. In these cases, the $\tau$ candidates are 
added to the signal template. 
The efficiency for electrons and muons misidentified as $\tau$ candidates 
is determined
by studying $Z\rightarrow \ell^+\ell^-$ events. Based on these studies the estimated contribution  from class 3 
background to the signal template is 2.8\%. 
The total contribution from class 2 and class 3 backgrounds ($Z$+jets,  $t \bar{t}\rightarrow \ell \ell$+jets, single top and dibosons) to the signal template is 15\%.
Table~\ref{tab:signal} shows the detailed composition of  the signal templates.
\begin{table}[!ht]
\begin{center}
\caption{\label{tab:signal} Composition of signal template: all events from MC simulation with a true $\tau$, $e$ or $\mu$ matched to the $\tau$ candidate. The number of events are normalized to the number expected from simulation. Regions 1 and 2 are $20\gev\le E_{\rm T}^{\tau} \le 35 \gev$ and  
$35\gev\le E_{\rm T}^{\tau} \le 100\gev$ respectively. The uncertainties represent the statistical uncertainties  of the MC samples.}
\begin{tabular}{l|cc}
\multicolumn{3}{c}{} \\ 
\hline\hline
Channel & Region 1 & Region 2\\
\hline 
$t\bar{t}\rightarrow\ell\tau_\mathrm{had}$+jets   &  611.5 $\pm$ 5.4  &   621.4 $\pm$ 5.4 \\
$t\bar{t}\rightarrow \ell\ell$+jets   &  \phantom{0}13.0 $\pm$ 0.7 &  \phantom{0}13.0 $\pm$ 0.7  \\
$Z$ $+$ jets   &  \phantom{0}54.5 $\pm$ 3.3 &  \phantom{0}45.3 $\pm$ 3.0 \\
Single top   &  \phantom{0}23.6 $\pm$ 2.3 &  \phantom{0}27.1 $\pm$ 2.4 \\
Dibosons   &  \phantom{00}1.5 $\pm$ 0.2 &  \phantom{00}2.2 $\pm$ 0.3  \\
Total   &  705.2 $\pm$ 6.8  &  709.5 $\pm$ 6.8 \\
\hline\hline

\end{tabular}

\end{center}
\end{table}

With these background templates and MC signal template ($S$),
a $\chi^2$ fit is performed with parameters to set the normalization of each template: $a \cdot \text{OS1} + b \cdot G+ c \cdot S$. 
The combined $e$ and $\mu$ channel results are obtained
by fitting to the sum of the distributions. Comparisons of the 
template shapes  of the $e$ and $\mu$ channel show they are identical within the uncertainties.

Two different $E_{\rm T}$ regions, $20\gev\le E_{\rm T}^{\tau} \le 35 \gev$ and  
$35\gev\le E_{\rm T}^{\tau} \le 100\gev$, are chosen 
such that each region has the same number of expected signal events.
Three parameters are used to fit both regions simultaneously: 
the fraction of $\tau$ candidates
in each $E_{\rm T}$ region that are gluon jets
and the total fraction of signal. In the fit the sum of signal and background must add 
up to the number of observed events in each $E_{\rm T}$ region and the amount of signal
 in the two regions is constrained by the ratio predicted from MC simulation.

%%%%%%%%%%%%%%%%%%%%%%%%%%%%%%%%%%%%%%%%%%%%%%%%%%%%%%%%%%%%%%%%%%%%%%%%
\subsection{Fit results}
\label{sec:ltaufits}

\begin{figure}[!hbt]
\begin{center}
\begin{tabular}{cc}
\epsfig{figure=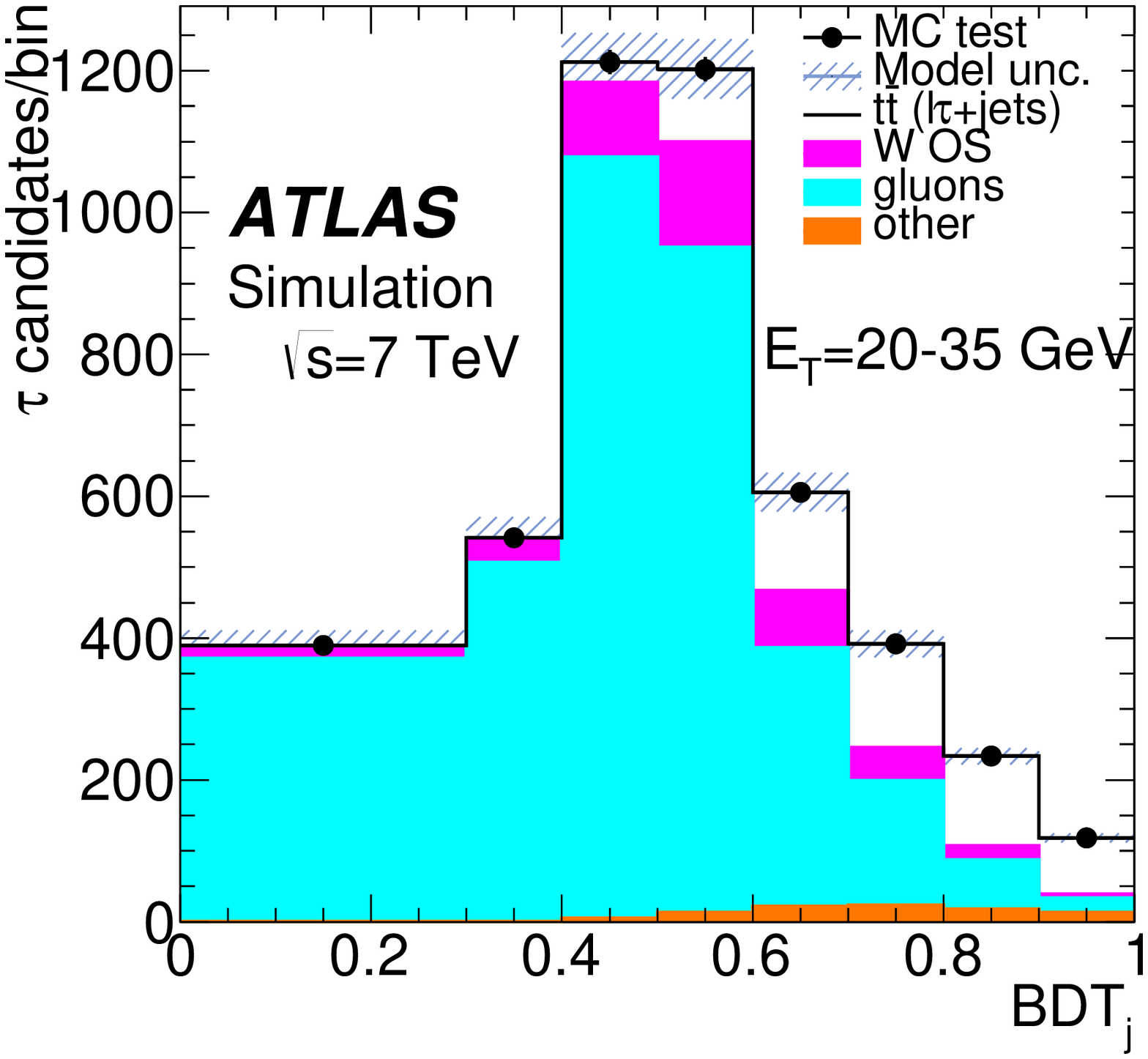,width=7.5cm} &
\epsfig{figure=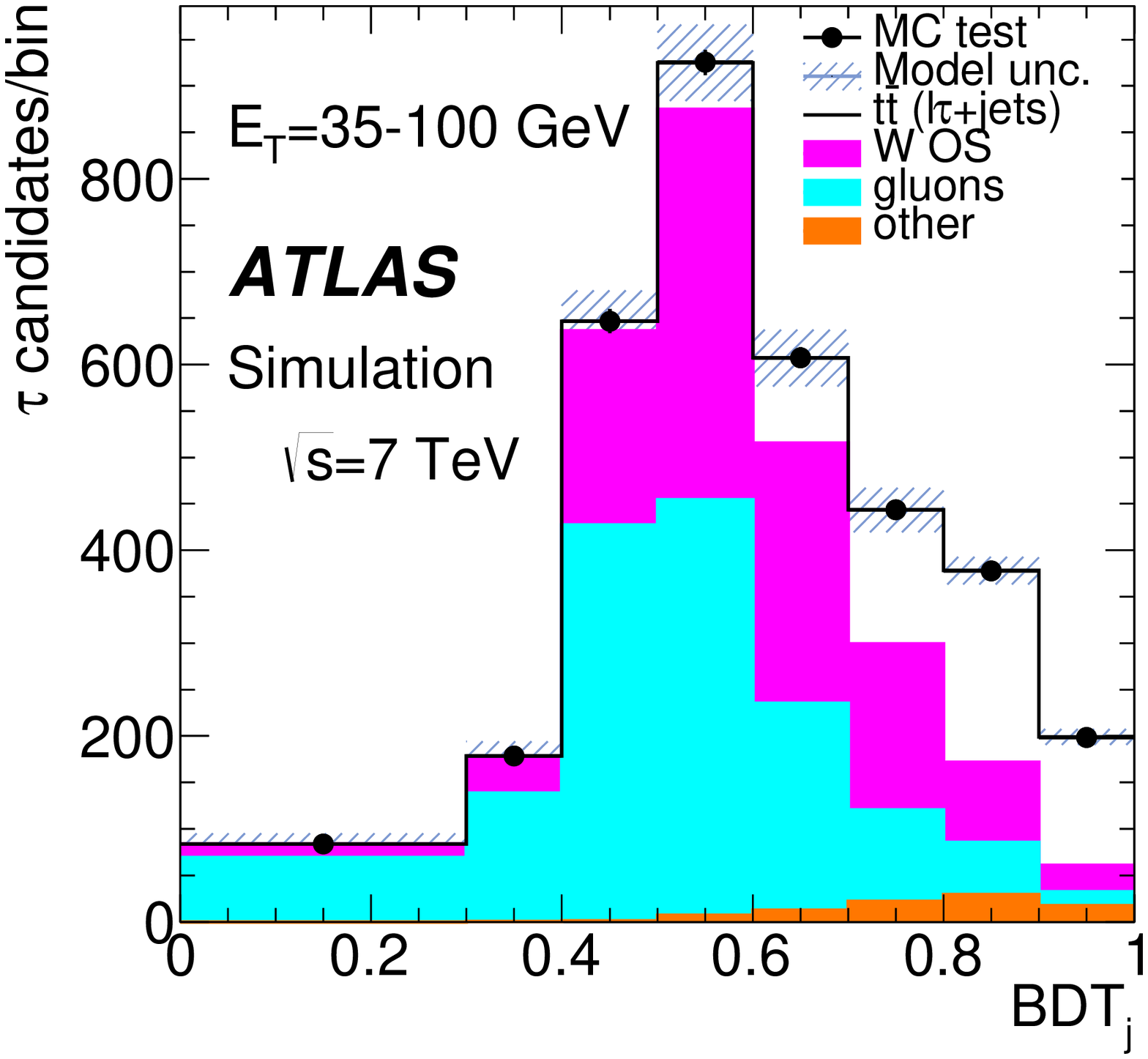,width=7.5cm} \\
(a) $20\gev\le E_{\rm T} \le 35 \gev$ & (b) $35\gev\le E_{\rm T} \le 100 \gev$\\
\end{tabular}
\caption{\label{fig:CorrMCfits} Fitted distributions of the $\tau$-jet discriminant BDT$_j$ MC using corrected 
background templates for two $E_T$ regions. The model uncertainty is the uncertainty of the templates used in the fits to the data.}
\end{center}
\end{figure}

The three-parameter fit was applied to MC samples to establish whether it can
extract the known signal without bias. The MC samples are made with events
from $t \overline{t}$, $W$+jets, $Z$+jets, single top and diboson final states 
satisfying the data selection criteria. The MC samples were split into two, one used as
the data to fit and the other to generate the templates for the fit. Figure~\ref{fig:CorrMCfits} shows 
these MC fit results  
after correcting the background templates derived from $W$+jets to account for
the different $E_{\rm T}$ distribution of the $\tau$ candidates
in the expected background to 
$t \overline{t}\rightarrow\ell\tau_\mathrm{had}$+jets. The model uncertainty shown in Figure~\ref{fig:CorrMCfits}
corresponds to the uncertainty of the templates in the fits to the data and used for ensemble tests.
The ensemble tests show that no bias is introduced by the fitting procedure.
The $\mu$ and $e$ channels are combined by adding together the distributions 
of both channels. The data BDT$_j$ distributions can have multiple
entries for an event as all $\tau$ candidates are considered. This has no impact on the $t \bar t\rightarrow \ell\tau_\mathrm{had}$+jets signal as there is only one $\tau$ lepton decaying to hadrons in that channel.

\begin{figure}[!hbt]
\begin{center}
\begin{tabular}{cc}
\epsfig{figure=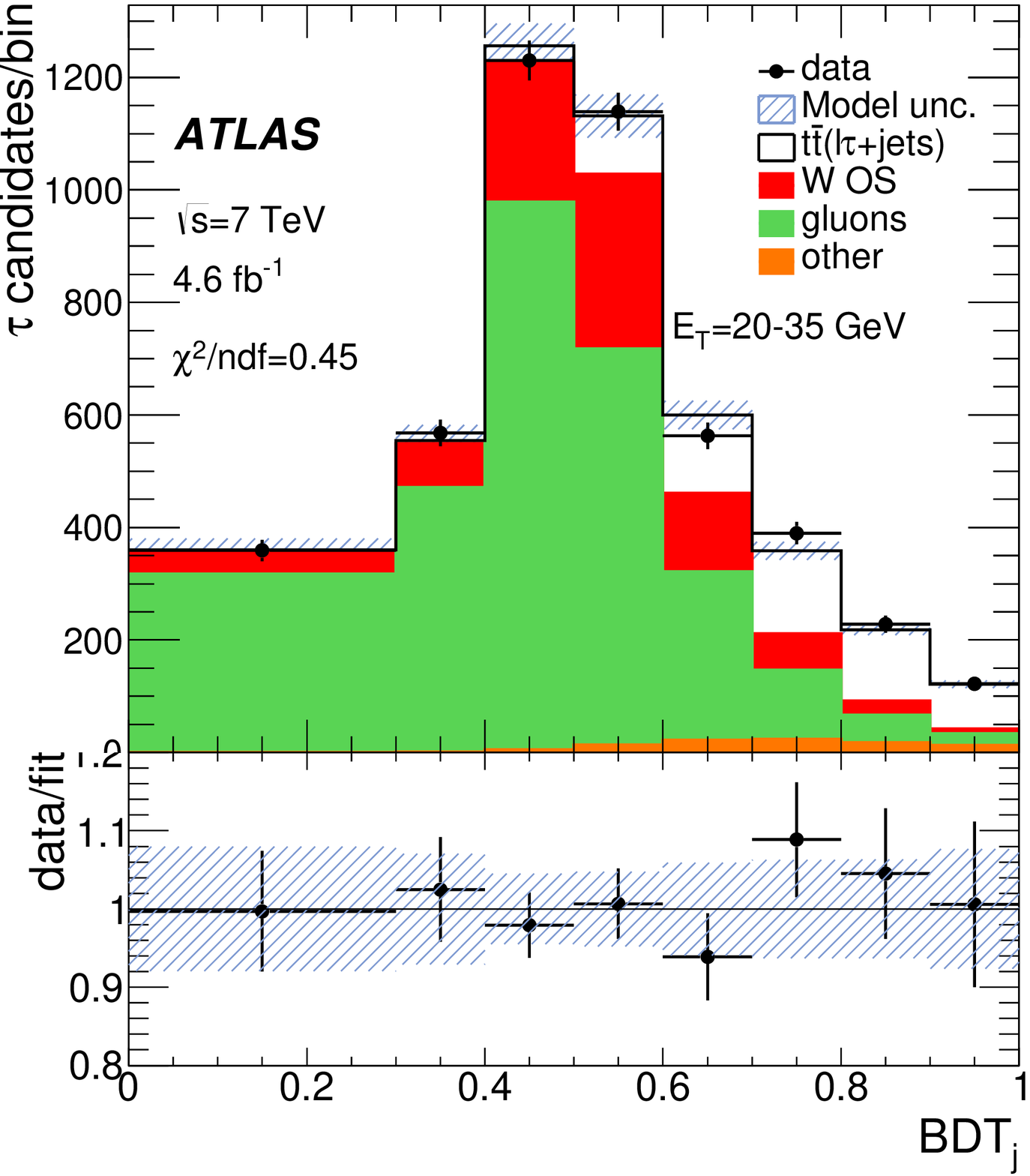,width=7.5cm} &
\epsfig{figure=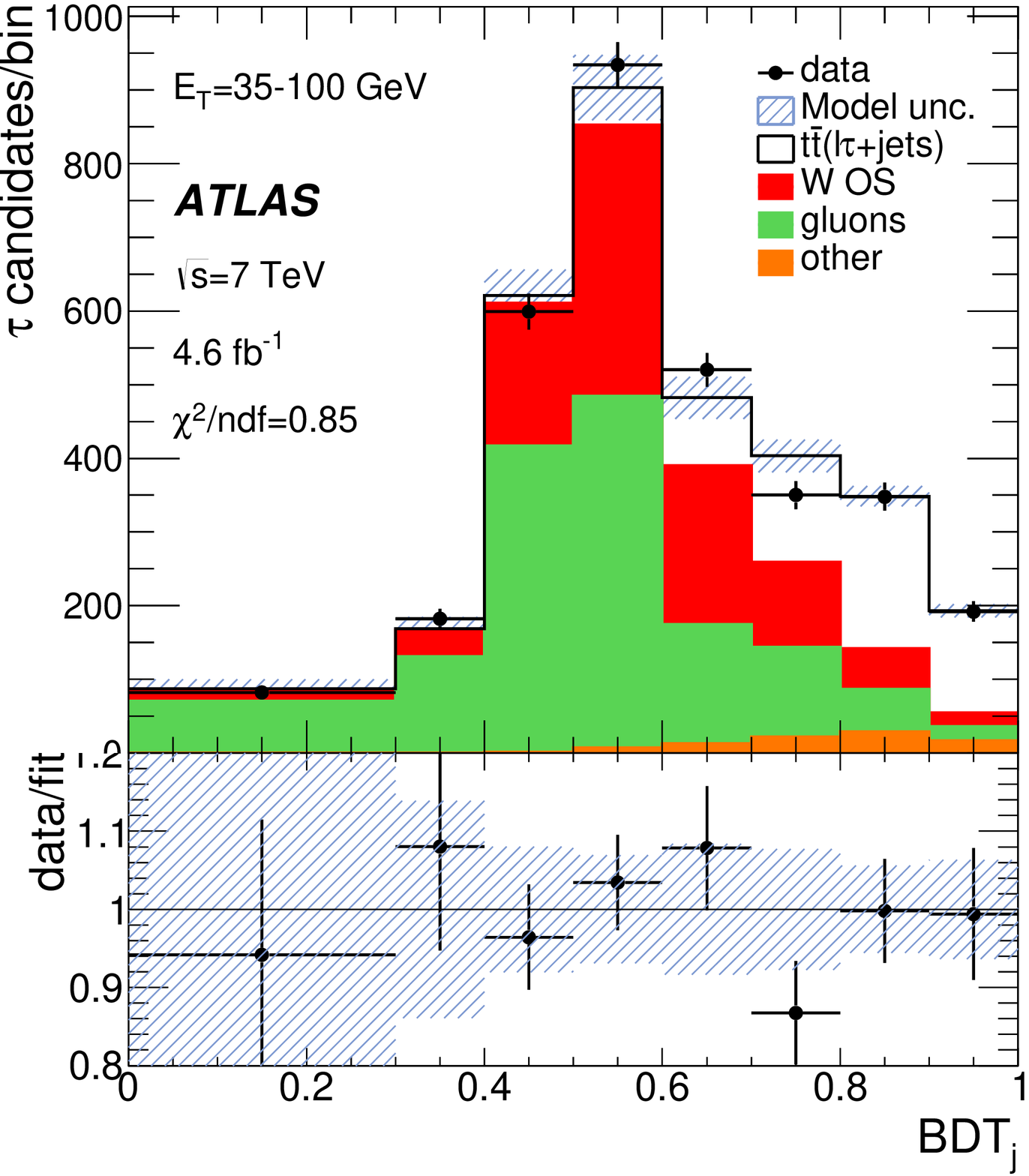,width=7.5cm} \\
(a) $20\gev\le E_{\rm T} \le 35 \gev$ & (b) $35\gev\le E_{\rm T} \le 100 \gev$\\
\end{tabular}
\caption{\label{fig:ltaufits} Fitted distributions of the $\tau$-jet discriminant BDT$_j$ in data using corrected 
background templates for (a) $20\gev\le E_{\rm T} \le 35 \gev$ and (b) $35\gev\le E_{\rm T} \le 100 \gev$.   The model uncertainty is the statistical uncertainty of the templates used in the fits.}
\end{center}
\end{figure}

The results of fitting the data are summarized in Table~\ref{tab:ltaufits}. $N^{\rm Fitted}_S$ is the number of 
signal template events. $N^{\rm Fitted}_{t \overline{t}}$ is the number of observed $t \bar t\rightarrow \ell\tau$+jets
events, obtained by subtracting the contributions from class 2 and class 3 
backgrounds (see Sec.~\ref{sec:fits}) from
$N^{\rm Fitted}_S$. The number of expected ($N^{\rm MC}_{t \overline{t}}$) is in good agreement with $N^{\rm Fitted}_{t \overline{t}}$. 
Figure~\ref{fig:ltaufits} shows the final results using these $\mu$ and $e$ channel combined templates. 

\begin{table}[h!]
\begin{center}
\caption{\label{tab:ltaufits} Numbers of events expected from MC simulation and fit results to the BDT$_j$ distribution using
background and signal templates as described in Sec.~\ref{sec:taubackground}.
$N^{\rm MC}_{t \overline{t}}$ is the expected number of 
$t \overline{t}\rightarrow\ell\tau_\mathrm{had}$+jets events for a cross section of 177 pb.
   B$_{{\rm non}~t\overline{t}~{\tau}}$ is the number of $\tau$ leptons
expected from sources other than $t \overline{t}\rightarrow\ell\tau_\mathrm{had}$+jets. B$_{\rm lepton}$ is the expected number of leptons
misidentified as $\tau$ leptons. $N^{\rm Fitted}_S$ is the number of events extracted with the signal template ($S$, see text) and  $N^{\rm Fitted}_{t \overline{t}}$= $N^{\rm Fitted}_S$-B$_{{{\rm non}~t\overline{t}~{\tau}}_{~}}$-B$_{{\rm lepton}_{~}}$} 
\begin{tabular}{l|ccc|cc}
\multicolumn{6}{c}{} \\  
\hline
\hline
 &$N^{\rm MC}_{{t \overline{t}}_{~}}$   & B$_{{{\rm non}~t\overline{t}~{\tau}}_{~}}$  & B$_{{\rm lepton}_{~}}$ & $N^{\rm Fitted}_{S_{~}}$ & $N^{\rm Fitted}_{{t \overline{t}}_{~}}$\\ 
\hline
$20<E^{\tau}_{\rm T}<35$ GeV    &  \phantom{0}611 $\pm$ 5 & \phantom{0}76.2 $\pm$ 3.5  & 17.1 $\pm$ 1.1 & N/A & N/A \\ \hline
$35<E^{\tau}_{\rm T}<100$ GeV   &  \phantom{0}621 $\pm$ 5 & \phantom{0}69.5 $\pm$ 3.3  & 17.6 $\pm$ 1.1 & N/A & N/A \\
\hline
\rule{0pt}{12pt}Combined $E^{\tau}_{\rm T}$ bins  &  1232 $\pm$ 8               & 146 $\pm$ 5 & 34.8 $\pm$ 1.5 & 1460 $\pm$ 60 (${\chi}^{2}/{\rm ndf}=$0.69) & 1280 $\pm$ 60 \\ \hline\hline
\end{tabular}

\end{center}
\end{table}

\begin{figure}
\begin{tabular}{cc}
\epsfig{figure=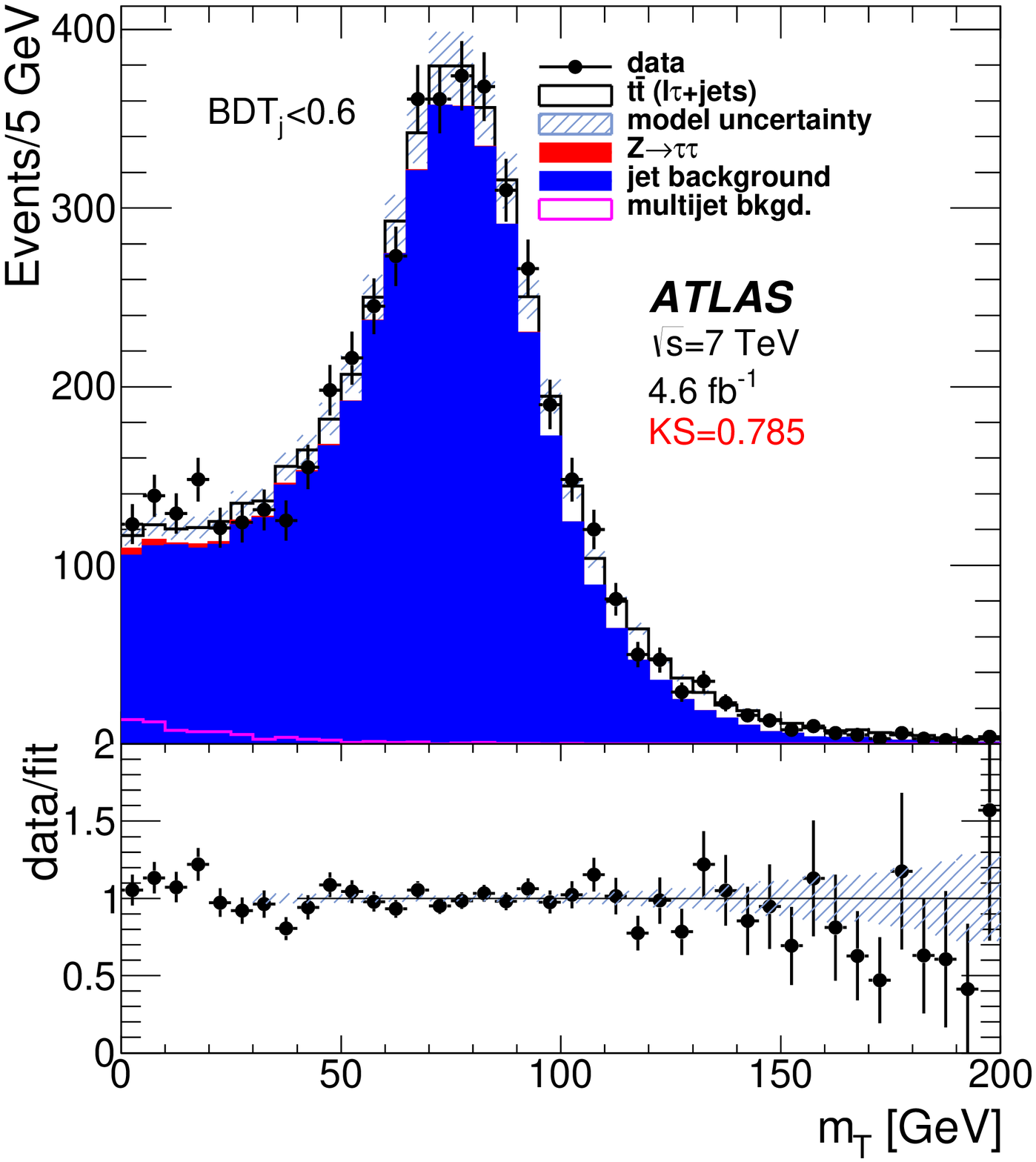,width=7.5cm} & 
\epsfig{figure=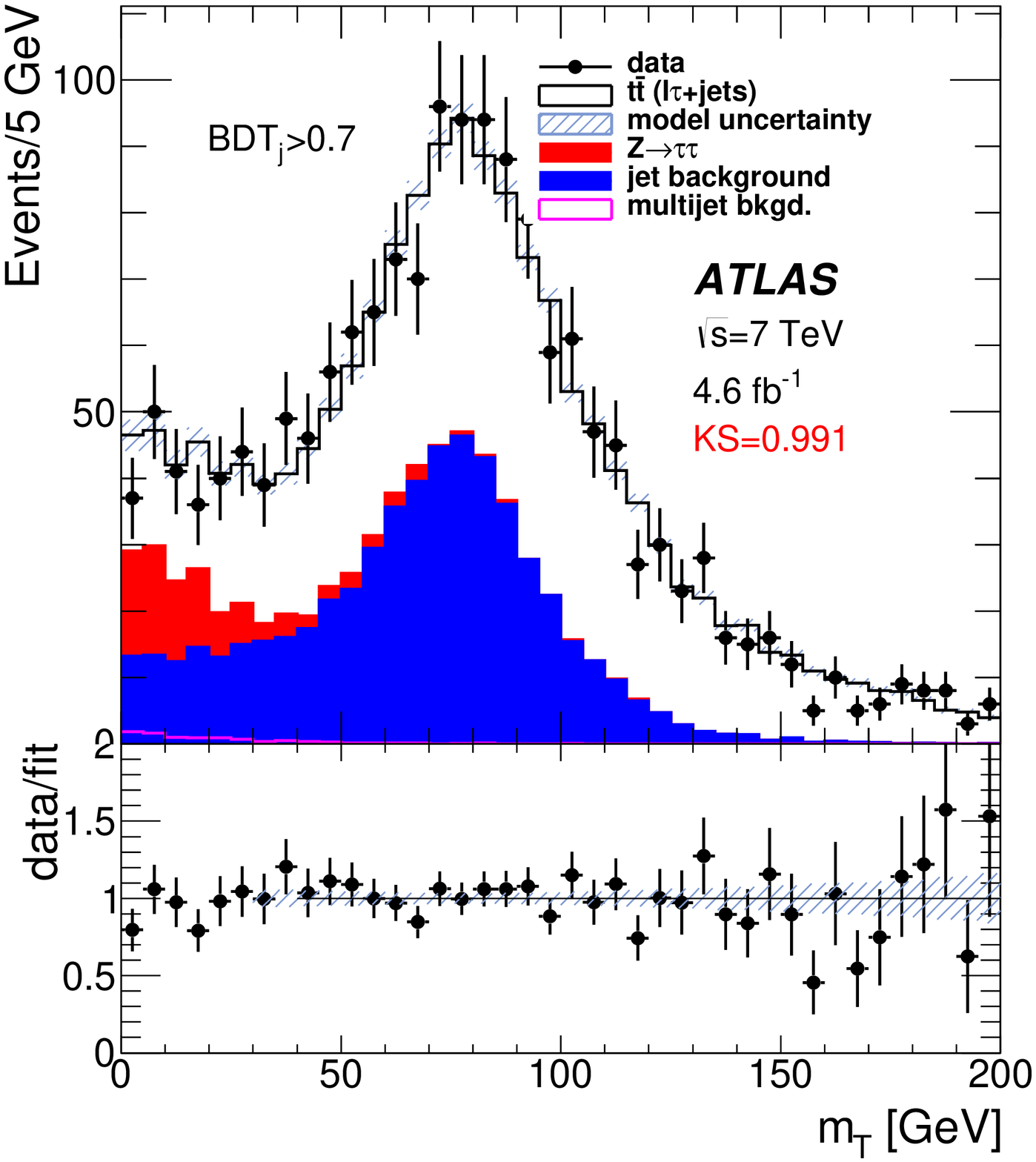,width=7.5cm} \\
(a) Background region, BDT$_j<$0.6. & (b) Signal region, BDT$_j>$0.7. \\
\end{tabular}
\caption{\label{fig:Mt_ltau} Transverse mass 
 distributions ($m_{\rm T}$) of $t \overline{t}\rightarrow\ell\tau_\mathrm{had}$+jets events.
The black points are data, the solid histograms the prediction based on the fits 
to the BDT$_j$ distributions. The jet background is the sum of all 
channels with jets misidentified as $\tau$ candidates normalized to the amount 
obtained from the fits to BDT$_j$ distributions. 
The multijet background is the estimated contribution from non-$t \overline{t}$ 
multijet processes and is included in the jet background. The model uncertainty is the statistical uncertainty of the templates used in the fits. KS is the value of the Kolmogorov-Smirnov goodness-of-fit test.}
\end{figure}

Jets misidentified as $\tau$ leptons come mostly from
$t \overline{t}\rightarrow \ell$+jets and from $W$+jets. Thus the $m_{\rm T}$ distributions
should show a Jacobian peak from a $W$ decay. The $t \overline{t}\rightarrow \ell\tau_\mathrm{had}$+jets events
have additional neutrinos, which produce a broader $m_{\rm T}$ distribution. 
 Figure \ref{fig:Mt_ltau}  shows the distributions from events
selected with BDT$_j<0.6$, which are mostly background, and for events selected
with BDT$_j>0.7$ where the ratio of signal to all background is 2:1. The plots include 
the predicted
distributions using the normalizations based on the fits to 
the BDT$_j$ distributions. 
The amount of $Z\rightarrow \tau \tau$ is normalized to the MC prediction. The data
are well reproduced in all cases.

\section{Measuring Cross Section and Branching Ratios}
\label{sec:BR}
In the SM 100\% of the top quark decays have one $W$ boson and a quark.
Therefore the top quark branching ratios into channels with leptons and jets are determined by
the $W$ decay branching ratios that have been measured with 0.3\% precision (assuming lepton universality)~\cite{Wbr} and are predicted
by the SM with an uncertainty of order 0.1\%. It is possible to derive the branching
ratios into all decay modes using the number of $t \overline{t}$ events extracted
in the previous sections assuming that the top quark branching ratios to leptons
and jets add up to 100\%. Any deviation from the $W$ branching ratios 
would be an indication of some process not predicted by the SM.
The following observed quantities are defined (where ${\cal A}_{ch}\cdot\epsilon_{ch}$ is the 
geometric detector acceptance times the efficiency of channel $ch$): 
\begin{itemize}
\item $N_{\mu j}$=(observed number of $t \overline{t}\rightarrow \mu$+jets)$/{\cal A}_{\mu j}\cdot\epsilon_{\mu j}$,
\item $N_{ej}$=(observed number of $t \overline{t}\rightarrow e$+jets)$/{\cal A}_{ej}\cdot\epsilon_{ej}$,
\item $N_{\mu\mu}$=(observed number of $t \overline{t}\rightarrow \mu +\mu$+jets)
/${\cal A}_{\mu \mu}\cdot\epsilon_{\mu \mu}$,
\item $N_{ee}$=(observed number of $t \overline{t}\rightarrow e+e$+jets)
/${\cal A}_{ee}\cdot\epsilon_{ee}$,
\item $N_{e\mu}$=(observed number of $t \overline{t}\rightarrow e+ \mu$+jets)
/${\cal A}_{e\mu}\cdot\epsilon_{e \mu}$, 
\item $N_{\ell\tau}$=(observed number of $t \overline{t}\rightarrow \ell+ \tau_{\mathrm had}$+jets)
/${\cal A}_{\ell\tau}\cdot\epsilon_{\ell \tau}$,
\item $N_{\ell j}$=$N_{\mu j}$+$N_{ej}$,
\item $N_{\ell\ell}$=$N_{\mu \mu}$+$N_{ee}$+$N_{e\mu}$. 
\end{itemize}
The following notation is used for the top quark branching ratios:
\begin{itemize}
\item $B_{\mu}$: top quark branching ratio to $\mu \nu_{\mu} (\nu_{\tau}) +X$,
\item $B_e$: top quark branching ratio to $e \nu_e (\nu_{\tau})+X$, 
\item $B_{\tau}$: top quark branching ratio to $\tau \nu_{\tau}+X$, with the $\tau$ lepton
decaying hadronically
\item $B_j$: top quark branching ratio to jets,
\item $B_{\ell}$: $B_\mu$+$B_e$.
\end{itemize}
The branching ratios $B_\mu$ and $B_e$ include events with leptonic $\tau$ decays.

With these definitions the following relations hold:
\begin{equation}
N_{\ell j}=2\sigma_{t \overline{t}}\cdot B_{\ell} \cdot B_j \cdot \cal L,
\end{equation}
\begin{equation}
N_{\ell\ell}=\sigma_{t \overline{t}}\cdot B_{\ell}^2 \cdot \cal L,
\end{equation}
\begin{equation}
N_{\ell\tau}=2\sigma_{t \overline{t}}\cdot B_{\ell} \cdot B_{\tau} \cdot \cal L,
\end{equation}
\begin{equation}
B_j+B_\ell+B_\tau=1,
\end{equation}
where $\sigma_{t \overline{t}}$ is the cross section for $t \overline{t}$ pair production and $\cal L$
is the integrated luminosity.
These four equations with four unknowns can be solved to obtain:
\begin{equation}
B_j=N_{\ell j}/(N_{\ell j}+2N_{\ell\ell}+N_{\ell\tau}),
\end{equation}
\begin{equation}
B_{\ell}=2N_{\ell\ell}/(N_{\ell j}+2N_{\ell\ell}+N_{\ell\tau}),
\end{equation}
\begin{equation}
B_{\tau}=N_{\ell\tau}/(N_{\ell j}+2N_{\ell\ell}+N_{\ell\tau}),
\end{equation}
\begin{equation}
\sigma_{t \overline{t}} \cdot {\cal L}=(N_{\ell j}+2N_{\ell\ell}+N_{\ell\tau})^2/4N_{\ell\ell}.
\end{equation}

From the numbers of $t \overline{t}$ events given in Tables~\ref{tab:ljet_fits}--
\ref{tab:ltaufits} and the acceptances given in Table~\ref{tab:accept} 
the values are obtained for $N_{\ell x}$ and given in Table~\ref{tab:N}. The $N_{\ell x}$ are in
units of events/pb$^{-1}$.
  
 After solving for $B_{\ell}$ one can solve for
$B_e$ and $B_\mu$  using ratios in the dilepton and the single-lepton channel:

\begin{equation}
  B_{\mu(e)}=2N_{\mu \mu(ee)}\cdot B_j/N_{\mu(e)j}\equiv a,
\end{equation}
\begin{equation}
B_{\mu(e)}=B_{\ell}\cdot \sqrt{N_{\mu\mu(ee)}/N_{\ell\ell}}\equiv b.
\end{equation}
The best values are obtained by minimizing
\begin{equation}
\chi^2=([B_{\mu(e)}-a]/\delta a)^2+([B_{\mu(e)}-b]/\delta b)^2,
\end{equation}
where $\delta a$ and $\delta b$ are the $a$ and $b$ uncertainties.
\begin{table}[h!bt]
\centering
\caption{\label{tab:accept} The acceptance $\times$ efficiency (${\cal A}_{ch}\cdot\epsilon_{ch}$) of each channel used to extract the number
of $t \overline{t}$ events after all selections. The ${\cal A}_{ch}\cdot\epsilon_{ch}$ are calculated by taking the
ratio of fully reconstructed MC events to MC generated events. The uncertainties represent the statistical uncertainties of the MC samples.}
\begin{tabular}{c|cccccc}
\multicolumn{6}{c}{} \\ 
\hline
\hline 
 & $e$+jets & $\mu$+jets & $ee$+jets & $\mu\mu$+jets& $e\mu$+jets &$\ell\tau$+jets \\
\hline
${\cal A}_{ch}\cdot\epsilon_{ch}$(\%) & 14.02$\pm$0.02 & 17.88$\pm$0.02 & 7.09$\pm$0.04 & 19.74$\pm$0.08 & 9.50$\pm$0.04 & 4.36$\pm$0.02 \\
\hline
\hline
\end{tabular}

\end{table}

\begin{table}[h!bt]
\centering
\caption{\label{tab:N} Measured number of events/pb$^{-1}$ for each channel and the
number predicted by the SM. Data uncertainties are statistical only. The SM uncertainty is calculated using the theoretical uncertainty of the NNLO+NNLL calculation of the 
cross section.}
\begin{tabular}{l|cc|ccc|c}
\multicolumn{7}{c}{} \\ 
\hline
\hline 
~&$N_{ej}$ &$N_{\mu j}$&$N_{ee}$ & $N_{\mu\mu}$& $N_{e\mu}$ &$N_{\ell\tau}$ \\
~&\multicolumn{2}{|c|}{$N_{\ell j}$} &\multicolumn{3}{|c|}{$N_{\ell\ell}$} & \\
\hline
Measured& 30.62$\pm$0.26&30.57$\pm$0.29&3.06$\pm$0.12&\phantom{0}3.19$\pm$0.10&6.06$\pm$0.12&6.39$\pm$0.30\\
~&\multicolumn{2}{|c|}{61.19$\pm$0.40}&\multicolumn{3}{|c|}{12.31$\pm$0.20}& \\
\hline
SM &30.40$\pm$1.2&30.40$\pm$1.2&2.86$\pm$0.11&\phantom{0}2.86$\pm$0.11&5.72$\pm$0.20&6.39$\pm$0.25\\ 
~& \multicolumn{2}{|c|}{60.64$\pm$2.4}&\multicolumn{3}{|c|}{10.95$\pm$0.44}& \\
\hline
\hline
\end{tabular}

\end{table}

\section{Systematic Uncertainties}

\label{sec:systematics}
Several sources of experimental and theoretical systematic  uncertainty are considered.
Lepton trigger, reconstruction and selection efficiencies are assessed in data and MC 
 simulation by comparing the
$Z \rightarrow\ell^{+}\ell^{-}$ events selected with the same object criteria
as used for the $t \overline{t}$ analyses.
Scale factors are applied to MC samples when calculating acceptances to account for any differences 
between predicted and observed efficiencies. The scale factors are evaluated by comparing the 
observed efficiencies with those determined  with simulated $Z$ boson events. 
Systematic uncertainties on these scale factors are evaluated by varying 
the selection of events used in the efficiency measurements and by 
checking the 
stability of the measurements over the course of data taking.	
The modeling of the lepton momentum scale and resolution is studied with 
reconstructed invariant mass distributions of $Z \rightarrow \ell^{+}\ell^{-}$\ candidate 
events, and these distributions are used to adjust the simulation accordingly~\cite{muons,electrons}.
	
The jet energy scale (JES), jet energy resolution (JER),
and their uncertainties  are derived by combining information from test-beam data, 
LHC collision data and simulation.
For jets within the acceptance, the JES uncertainty varies
in the range 4\%--8\% as a function of jet $\pt$ and $\eta$~\cite{JES}. The
$b$-tagging efficiency and its uncertainty is determined using
a sample of jets containing muons~\cite{MV1}.
The effect of all these variations on the final result is evaluated by varying each source of 
systematic uncertainty by $\pm 1\sigma$ in the  MC-derived templates and 
fitting all the distributions with the new templates.

The uncertainty in the kinematic distributions of the $t \overline{t}$ signal events
gives rise to systematic uncertainties in the signal acceptance, with contributions 
from the choice of generator, the modeling of initial- and final-state
radiation (ISR/FSR) and the choice of PDF set.
The generator uncertainty is evaluated by comparing the  {\sc MC@NLO} and 
ALPGEN~\cite{Alpgen} predictions  with those of
POWHEG~\cite{powheg2} interfaced to either HERWIG or PYTHIA.
The PDF uncertainty is evaluated following the PDF4LHC recommendation~\cite{PDF4LHC}. 
An event-by-event weighting is applied to a
default MC@NLO sample that uses the central value of CT10~\cite{CT10}. 
MSTW2008 ~\cite{MSTWPDF} and 
NNPDF2.0 ~\cite{NNPDF1, NNPDF2} sets are taken to estimate the systematic uncertainty due to the PDF. 
The uncertainty due to ISR/FSR is evaluated using the ALPGEN generator 
interfaced to the PYTHIA shower model, and by varying the parameters controlling ISR
and FSR in a range consistent with experimental data~\cite{ISR/FSR}.
The dominant uncertainty in this category of systematic uncertainties is 
the modeling of ISR/FSR. In addition there is an uncertainty in the $W$+jets
MC simulation due to the uncertainty in the heavy flavor component of the jets. The systematic uncertainty
from single top MC simulation has a negligible impact on the overall systematic uncertainty. 

The $\tau$ identification uncertainty is derived from a template fit to the BDT$_j$ distribution
from an enriched $Z \rightarrow \tau^{+} \tau^{-}$ data sample selected 
with the same $\mu$ and $\tau$ 
candidate requirements as the sample for this analysis, but with fewer than two jets and 
$m_{\rm T}<20\gev$ to remove $W+$jets events. The background templates are 
the $W$+1-jet OS and the gluon template used in the fit to the $t \bar t$ data sample. 
The signal template is the BDT$_j$ distribution from $Z\rightarrow \tau^{+} \tau^{-}$ MC events.
The uncertainty includes the statistical uncertainty of the data samples, the uncertainty
in the $Z$ inclusive cross section measured by ATLAS~\cite{ZXsec} (excluding luminosity uncertainty)  and jet energy
scale uncertainty. The signal template shape uncertainty, estimated 
from fits to the $Z \rightarrow \tau^{+} \tau^{-}$ data sample, is found to be negligible. The uncertainty on the number of misidentified electrons ($<0.5$\%), determined from an enriched $Z\rightarrow e^{+}e^{-}$ data sample, is included. In addition there is an uncertainty in 
the correction applied to the $\tau$ background templates derived from $W$+jets data
to account for the different $E_{\rm T}$ distribution of the $\tau$ candidates in the 
expected background to 
$t \overline{t}\rightarrow\ell\tau_\mathrm{had}$+jets.

The calculated systematic uncertainties for the inclusive cross section measured with the $\ell\tau_\mathrm{had}$+jet channel are given in Table~\ref{tab:fits-systematics-combpt}. Table~\ref{tab:sys1} gives the systematic uncertainties estimated
when combining all channels. The  uncertainty on the measured integrated luminosity is estimated to be 1.8\% \cite{luminew}. As expected the systematic uncertainties are substantially larger in the measurement of the cross section based on the $\ell\tau_\mathrm{had}$+jets channel alone than in the combination of all channels. 
The largest uncertainty in the combined cross-section measurement and in the branching ratio measurements 
is due to the JES uncertainty, followed by the MC generator and the uncertainty in the heavy-flavor 
component of $W$+jets. The uncertainties on the
measured branching ratios are significantly smaller than on the measured inclusive cross section,
 as expected due to cancellations. $B_{\tau}$ has a larger systematic 
uncertainty than the other branching ratios due to uncertainties on $\tau$ 
identification that do not cancel in the ratios.

\begin{table}[H]
\begin{center}
\caption{\label{tab:fits-systematics-combpt} Absolute systematic uncertainties, in pb, for the cross-section measurements
with the $t \overline{t}\rightarrow \ell\tau_\mathrm{had}$+jets channel. The $e$ and $\mu$ uncertainties are the sum in quadrature
of trigger, reconstruction and selection efficiency uncertainties. The $\tau$ identification uncertainty includes electrons misidentified as $\tau$ leptons. }
\begin{tabular}{l|c}
\multicolumn{2}{c}{} \\   
\hline\hline
 & Absolute uncertainties [pb]\\ \hline
$\mu$ uncertainty               & 1.7   \\
$e$ uncertainty                &  3.0  \\
Jet energy scale          & $-5.5$ / $+6.8$ \\
Jet energy resolution           & 1.5         \\
ISR/FSR              & 12.3 \\
MC generator         & 10.1 \\
PDF         & 0.6      \\
$b$-tag              & $-8.3$ /  $ +10.0$  \\
$\tau$ identification         & 8.0  \\
$\tau$ background correction & 5.6 \\
\hline
Total                & $-22$/$+23$\\
\hline
Luminosity & 3.3 \\
\hline
\hline
\end{tabular}

\end{center}
\end{table}

\begin{table}[H]
\begin{center}
\caption{\label{tab:sys1}Relative systematic uncertainties (\%) for cross section and 
branching ratio measurements. 
The systematic uncertainties for $B_e$ and $B_{\mu}$ 
(not shown) are 100\% correlated with the $B_\ell$ uncertainties and 
of the same size. The $e$ and $\mu$ uncertainties are the sum in quadrature
of trigger, reconstruction and selection efficiency uncertainties. 
The MC generator uncertainty  
is the difference between POWHEG interfaced with PYTHIA and ALPGEN interfaced 
with HERWIG. HF stands for heavy-flavor.}
\begin{tabular}{l|cccc}
\multicolumn{5}{c}{} \\ 
\hline
\hline
~ &$\sigma_{t \overline{t}}$&$B_{j}$&$B_{\ell}$& $B_{\tau}$ \\
\hline
$\mu$ uncertainty & 1.3  & 0.15 & 0.6 & 0.5 \\
$e$ uncertainty  & 1.1  & 0.15 & 0.5 & 0.5 \\
Jet energy scale  & $-6.9/+4.9$ & $-1.6/+1.4$ & $-1.9/+2.7$ & $-3.8/+4.3$ \\
Jet energy resolution            & 1.2 & 0.3 & 0.8 &0.7 \\
ISR/FSR    & 2.0 & 0.3 & 1.3 & 4.0 \\
MC generator       & 3.6    & 0.6 & 0.8 & 1.9 \\
PDF             & 2.9 & 0.3 & 0.1 & 0.3 \\
$b$-tag             & $-1.3/+5.0$ & 0.3 & 1.0 & 1.5 \\
$\tau$ identification         & 0.5  & 0.15& 1.1 & 3.5 \\
$\tau$ background correction & 0.2 & $<$0.1 & $<$0.1 & 2.5 \\
$W$+jets HF content   &  $-4.1/+2.7$ & $-1.0/+0.7$ & $-1.1/+2.3$ & $-1.3/+2.1$ \\
\hline
Total          &  $-9.7/+9.2$ & $-2.1/+1.8$ & $-3.4/+4.2$ & $-7.1/+7.6$  \\
\hline
Luminosity & 1.8 & $<$0.1 & $<$0.1 & $<$0.1 \\
\hline
\hline
\end{tabular}

\end{center}
\end{table}

\clearpage

\section{Results}
\label{sec:results}
The inclusive $t \overline t$ cross
section using only the $\ell\tau_\mathrm{had}$+jets channel is derived from the number of observed $t \overline{t}\rightarrow \ell\tau$+jets events given in Table~\ref{tab:ltaufits} (Sec.~\ref{sec:ltaufits}):

$$\sigma_{t \overline{t}}=183\pm9\stat \pm 23 \syst \pm 3~(\text{lumi.})~{\rm pb}$$
This result is consistent with the previous ATLAS 
measurement, 186$\pm$25 pb~\cite{PLB}.
This measurement differs from the earlier one in  that it uses only 
$\tau$s decaying into one charged hadron and a different background model to reduce the systematic uncertainties
in the branching ratios.
The results from combining all channels to extract the top quark branching ratios are given in Table~\ref{tab:final}.
The measured cross section of
$178\pm 17$ pb is 
in good agreement with those obtained by ATLAS for individual channels~\cite{ljetsPLB,dilJHEP,emub}.
The selection criteria for this measurement were optimized for
the $t \overline{t}\rightarrow \ell\tau_\mathrm{had}$+jets channel, which has the largest uncertainty,
and then applied uniformly to all channels, ensuring no event overlap between them to exploit 
cancellation of systematic uncertainties in the ratios.
This reduces the systematic uncertainties
in the branching ratio measurements but it is not optimal for 
a cross-section measurement combining all channels. The systematic uncertainty on the inclusive cross section obtained by combining 
the samples used
for this measurement is larger than the best ATLAS inclusive 
cross-section measurement~\cite{emub}, which achieved much smaller uncertainties because it was
designed to minimize the systematic uncertainties related to jets, including the b-tagging efficiency and the jet energy scale. All cross-section measurements are in 
good agreement 
with the NNLO+NNLL theoretical 
prediction $177.3\pm9.0^{+4.6}_{-6.0}$ pb (calculated for a top mass of $172.5\gev$~\cite{Czakon, Czakon2}). 

The branching ratios into leptons and jets are in good agreement with the SM prediction
that the top quark decays 100\% to $W$+quark. The precision of the measurements ranges from
2.3\% for $B_j$ to 7.6\% for $B_{\tau}$. The $B_e$ and $B_{\mu}$
include the leptonic decay of $\tau$ leptons while $B_{\tau}$
includes only the hadronic decays of $\tau$ leptons.
There is no evidence for any non-SM top quark decay or for any non-SM process contribution that could
affect these measurements.  For example, the measured branching ratio $B_\tau$ will vary by more
than the observed uncertainty if the branching ratio $\tilde{t}\rightarrow b \nu_{\tau} \tilde{\tau}$ times
the $\tilde{t}\tilde{\bar t}$ production cross section ($\sigma_{\tilde{t} \tilde{\overline{t}}}$)
is greater than 3\% of $\sigma_{t \overline{t}}$. The predicted $\sigma_{\tilde{t} \tilde{\overline{t}}}$
depends on $\tilde{t}$ mass ($m_{\tilde{t}}$); it is equal to $\sigma_{t \overline{t}}$ for
$m_{\tilde{t}}=120$ GeV and 12\% of $\sigma_{t \overline{t}}$ for $m_{\tilde{t}}=180$ GeV~\cite{Beenakker}.

\begin{table}[h!bt]
\centering
\caption{\label{tab:final}Measured cross section (pb) and top quark branching ratios (\%)
including statistical and systematic uncertainties without imposing lepton universality. The top quark branching ratios add up to 100.2\% because of rounding precision. The uncertainty on the SM prediction for the cross section is the uncertainty in the NNLO+NNLL theoretical calculation~\cite{Czakon,Czakon2}. 
The SM branching ratios are the predicted $W$ branching ratios. The LEP measurements represent the $W$ branching ratios
obtained by combining results for ALEPH, DELPHI, L3 and OPAL collaborations imposing lepton universality~\cite{Wbr}. The LEP entries $B_e$ and $B_{\mu}$ include the $\tau$ leptonic decays that have been subtracted from $B_{\tau}$.}
\begin{tabular}{l|c|c|c}
\multicolumn{4}{c}{} \\ 
\hline
\hline
~ & Measured & SM  & LEP \\
~ & (top quark) &  & ($W$) \\
\hline
$\sigma_{t \overline{t}}$ & $178\pm 3 \stat \pm 16 \syst \pm 3~\text{(lumi.)}$ pb
                          & $177.3\pm9.0^{+4.6}_{-6.0}$ pb & \\
\hline
$B_{j}$ & $66.5\pm 0.4 \phantom{0}\stat \pm 1.3 \syst$ & 67.51$\pm$0.07 &  67.48$\pm$0.28 \\
$B_{e}$ & $13.3 \pm 0.4 \phantom{0}\stat \pm 0.5 \syst$ & 12.72$\pm$0.01 & 12.70$\pm$0.20\\
$B_{\mu}$ & $13.4 \pm 0.3 \stat \pm 0.5 \syst$ & 12.72$\pm$0.01 & 12.60$\pm$0.18\\
$B_{\tau}$ & \phantom{0}$7.0 \pm 0.3 \stat \pm 0.5 \syst$ & \phantom{0}7.05$\pm$0.01 & \phantom{0}7.20$\pm$0.13\\
\hline
\hline
\end{tabular}

\end{table}

\section{Conclusion}
\label{sec:conclusion}
The inclusive cross section for producing $t \overline{t}$ pairs in $pp$ collisions at a center-of-mass energy of
$\sqrt{s}=7\tev$ at the LHC has been measured with the ATLAS detector and an integrated luminosity of 4.6 fb$^{-1}$ 
using the $\ell\tau_\mathrm{had}$+jets channel alone, as
 $\sigma_{t \overline{t}}~=183\pm23$ pb, and as a single parameter to fit the channels
$\ell$+jets, $\ell\ell$+jets and $\ell\tau_\mathrm{had}$+jets, to be $178\pm 17$ pb. 
These 
are in agreement with all other cross-section measurements obtained by ATLAS and CMS.
All cross-section measurements are fully compatible  
with the NNLO+NNLL theoretical 
prediction.
Top quark branching ratios have also been measured and found to be  in 
good agreement with branching ratios predicted by the SM.
 The precision ranges from 
2.3\% for the decays to jets to 7.6\% for the decays to $\tau\nu$+jet. 
There is no evidence for any non-SM
process affecting these branching ratios.
\newline
~

% Acknowledgements for papers with collision data
% Version 23-Mar-2015
We thank CERN for the very successful operation of the LHC, as well as the
support staff from our institutions without whom ATLAS could not be
operated efficiently.

We acknowledge the support of ANPCyT, Argentina; YerPhI, Armenia; ARC,
Australia; BMWFW and FWF, Austria; ANAS, Azerbaijan; SSTC, Belarus; CNPq and FAPESP,
Brazil; NSERC, NRC and CFI, Canada; CERN; CONICYT, Chile; CAS, MOST and NSFC,
China; COLCIENCIAS, Colombia; MSMT CR, MPO CR and VSC CR, Czech Republic;
DNRF, DNSRC and Lundbeck Foundation, Denmark; EPLANET, ERC and NSRF, European Union;
IN2P3-CNRS, CEA-DSM/IRFU, France; GNSF, Georgia; BMBF, DFG, HGF, MPG and AvH
Foundation, Germany; GSRT and NSRF, Greece; RGC, Hong Kong SAR, China; ISF, MINERVA, GIF, I-CORE and Benoziyo Center, Israel; INFN, Italy; MEXT and JSPS, Japan; CNRST, Morocco; FOM and NWO, Netherlands; BRF and RCN, Norway; MNiSW and NCN, Poland; GRICES and FCT, Portugal; MNE/IFA, Romania; MES of Russia and NRC KI, Russian Federation; JINR; MSTD,
Serbia; MSSR, Slovakia; ARRS and MIZ\v{S}, Slovenia; DST/NRF, South Africa;
MINECO, Spain; SRC and Wallenberg Foundation, Sweden; SER, SNSF and Cantons of
Bern and Geneva, Switzerland; NSC, Taiwan; TAEK, Turkey; STFC, the Royal
Society and Leverhulme Trust, United Kingdom; DOE and NSF, United States of
America.

The crucial computing support from all WLCG partners is acknowledged
gratefully, in particular from CERN and the ATLAS Tier-1 facilities at
TRIUMF (Canada), NDGF (Denmark, Norway, Sweden), CC-IN2P3 (France),
KIT/GridKA (Germany), INFN-CNAF (Italy), NL-T1 (Netherlands), PIC (Spain),
ASGC (Taiwan), RAL (UK) and BNL (USA) and in the Tier-2 facilities
worldwide.
%----------------------------------------------

\clearpage

%\bibliographystyle{atlasBibStyleWoTitle}
%\bibliography{topBR}
\providecommand{\href}[2]{#2}\begingroup\raggedright\endgroup

\newpage
% ATLAS Collaboration author list
% Data extracted on 11-Aug-2015 for paper reference TOPQ-2014-01
%\documentclass[11pt]{article}
%\usepackage{a4wide}\begin{document}
\begin{flushleft}
{\Large ATLAS Collaboration}

\bigskip

G.~Aad$^{\rm 85}$,
B.~Abbott$^{\rm 113}$,
J.~Abdallah$^{\rm 151}$,
O.~Abdinov$^{\rm 11}$,
R.~Aben$^{\rm 107}$,
M.~Abolins$^{\rm 90}$,
O.S.~AbouZeid$^{\rm 158}$,
H.~Abramowicz$^{\rm 153}$,
H.~Abreu$^{\rm 152}$,
R.~Abreu$^{\rm 30}$
% ...
\bigskip
\\
$^{1}$ Department of Physics, University of Adelaide, Adelaide, Australia\\
% ...
$^{11}$ Institute of Physics, Azerbaijan Academy of Sciences, Baku, Azerbaijan\\
% ...
$^{30}$ CERN, Geneva, Switzerland\\
% ...
$^{85}$ CPPM, Aix-Marseille Universit{\'e} and CNRS/IN2P3, Marseille, France\\
% ...
$^{90}$ Department of Physics and Astronomy, Michigan State University, East Lansing, Michigan, USA\\
% ...
$^{107}$ Nikhef National Institute for Subatomic Physics and University of Amsterdam, Amsterdam, Netherlands\\
% ...
$^{113}$ Homer L. Dodge Department of Physics and Astronomy, University of Oklahoma, Norman, oklahoma, USA\\
% ...
$^{151}$ Institute of Physics, Academia Sinica, Taipei, Taiwan\\
$^{152}$ Department of Physics, Technion: Israel Institute of Technology, Haifa, Israel\\
$^{153}$ Raymond and Beverly Sackler School of Physics and Astronomy, Tel Aviv University, Tel Aviv, Israel\\
% ...
$^{158}$ Department of Physics, University of Toronto, Toronto, Ontario, Canada\\
% ...
$^{s}$ Also at CERN, Geneva, Switzerland\\
$^{t}$ Also at Georgian Technical University (GTU),Tbilisi, Georgia\\
% ...
\end{flushleft}

%\end{document}
% Created with ./xml2latex.py
\end{document}